\documentclass[12pt]{article}
\usepackage{geometry} 
\usepackage{graphicx}
\usepackage{float}
\usepackage{subfig}
\usepackage{url}
\usepackage{mathtools}
\usepackage{color}
\usepackage[authoryear]{natbib}
\usepackage{fullpage}
\usepackage{setspace}
\usepackage{lineno}
\usepackage{titling}
\newcommand{\Exp}{\mathop{\mbox{Exp}}}
\newcommand{\cov}{\mathop{\mbox{Cov}}}

\title{Disentangling the effects of geographic and ecological isolation on genetic differentiation}
\date{\vspace{-5ex}}
\author{Gideon S. Bradburd$^{1,a}$, Peter L. Ralph$^{2,b}$, Graham M. Coop$^{1,c}$}

\begin{document}
\maketitle
\hspace{-0.25in}\textsuperscript{1}Center for Population Biology, Department of Evolution and Ecology, University of California, Davis, CA 95616\\
\textsuperscript{2}Department of Molecular and Computational Biology, University of Southern California, Los Angeles, CA 90089\\
\textsuperscript{a}gbradburd@ucdavis.edu; 
\textsuperscript{b}pralph@usc.edu;
\textsuperscript{c}gmcoop@ucdavis.edu

\newpage

\begin{abstract}
Populations can be genetically isolated both by geographic distance and by differences in their ecology or environment that decrease the rate of successful migration. Empirical studies often seek to investigate the relationship between genetic differentiation and some ecological variable(s) while accounting for geographic distance, but common approaches to this problem (such as the partial Mantel test) have a number of drawbacks. In this article, we present a Bayesian method that enables users to quantify the relative contributions of geographic distance and ecological distance to genetic differentiation between sampled populations or individuals.  We model the allele frequencies in a set of populations at a set of unlinked loci as spatially correlated Gaussian processes, in which the covariance structure is a decreasing function of both geographic and ecological distance. Parameters of the model are estimated using a Markov chain Monte Carlo algorithm.  We call this method Bayesian Estimation of Differentiation in Alleles by Spatial Structure and Local Ecology (\emph{BEDASSLE}), and have implemented it in a user-friendly format in the statistical platform R. We demonstrate its utility with a simulation study and empirical applications to human and teosinte datasets. 
\end{abstract}

\newpage
\section*{Introduction}

The level of genetic differentiation between populations is determined by the homogenizing action of gene flow balanced against differentiating processes such as local adaptation, different adaptive responses to shared environments, and random genetic drift.  Geography often limits dispersal, so that the rate of migration is higher between nearby populations and lower between more distant populations.  The combination of local genetic drift and distance-limited migration results in local differences in allele frequencies, the magnitude of which increases with geographic distance, resulting in a pattern of isolation by distance \citep{Wright1943}. Extensive theoretical work has described expected patterns of isolation by distance under a variety of models of genetic drift and migration \citep{Charlesworth2003} in both equilibrium populations in which migration and drift reach a balance, and under non-equilibrium demographic models, such as population expansion or various scenarios of colonization \citep{Slatkin1993}.  A range of theoretical approaches have been applied, with authors variously computing probabilities of identity of gene lineages \citep[e.g.][]{Malecot1975, Rousset1997} or correlations in allele frequencies \citep[e.g.][]{SlatkinMaruyama1975, WeirCockerham1984}, or working with the structured coalescent \citep[e.g.][]{Hey1991, NordborgKrone2002}.  Although these approaches differ somewhat in detail, their expectations can all be described by a pattern in which allele frequencies are more similar between nearby populations than between distant ones.

In addition to geographic distance, populations can also be isolated by ecological and environmental differences if processes such as dispersal limitations \citep{Wright1943}, biased dispersal \citep[e.g.][]{EdelaarBolnick2012}, or selection against migrants due to local adaptation \citep{Wright1943, Hendry2004} decrease the rate of successful migration. Thus, in an environmentally heterogeneous landscape, genome-wide differentiation may increase between populations as either geographic distance or ecological distance increase.  The relevant ecological distance may be distance along a single environmental axis, such as difference in average annual rainfall, or distance along a discrete axis describing some landscape or ecological feature not captured by pairwise geographic distance, such as being on serpentine versus non-serpentine soil, or being on different host plants.  

Isolation by distance has been observed in many species \citep{VekemansHardy2004, Meirmans2012}, with a large literature focusing on identifying other ecological and environmental correlates of genomic differentiation.  The goals of these empirical studies are generally 1) to determine whether an ecological factor is playing a role in generating the observed pattern of genetic differentiation between populations and, 2) if it is, to determine the strength of that factor relative to that of geographic distance. The vast majority of this work makes use of the partial Mantel test to assess the association between pairwise genetic distance and ecological distance while accounting for geographic distance \citep{Smouse1986}. 

A number of valid objections have been raised to the reliability and interpretability of the 
partial Mantel \citep[e.g.][]{LegendreFortin2010,Guillot2013}. 
First, because the test statistic of the Mantel test is a matrix correlation, it assumes a linear dependence between the distance variables, and will therefore behave poorly if there is a nonlinear relationship \citep{LegendreFortin2010}.  Second, the Mantel and partial Mantel tests can exhibit high false positive rates when the variables measured are spatially autocorrelated (e.g., when an environmental attribute, such as serpentine soil, is patchily distributed on a landscape), since this structure is not accommodated by the permutation procedure used to assess significance \citep{Guillot2013}. Finally, in our view the greatest limitation of the partial Mantel test in its application to landscape genetics may be that it is only able to answer the first question posed above --- whether an ecological factor plays a role in generating a pattern of genetic differentiation between populations --- rather than the first \emph{and} the second --- the strength of that factor relative to that of geographic distance.  By attempting to control for the effect of geographic distance with matrix regressions, the partial Mantel test makes it hard to simultaneously infer the effect sizes of geography and ecology on genetic differentiation, and because the correlation coefficients are inferred for the matrices of post-regression residuals, the inferred effects of both variables are not comparable --- they are not in a common currency.  
We perceive this to be a crucial lacuna in the populations genetics methods toolbox, as studies quantifying the effects of local adaptation \citep[e.g.][]{RosenblumHarmon2011}, host-associated differentiation \citep[e.g.][]{DresMallet2002, Gomez-Diaz2010}, or isolation over ecological distance \citep[e.g.][]{Andrew2012, Mosca2012} all require rigorous comparisons to the effect of isolation by geographic distance.  

In this article, we present a method that enables users to quantify the relative contributions of geographic distance and ecological distance to genetic differentiation between sampled populations or individuals.  To do this, we borrow tools from geostatistics \citep{Diggle1998} and model the allele frequencies at a set of unlinked loci as spatial Gaussian processes.   We use statistical machinery similar to that employed by the Smooth and Continuous AssignmenTs (SCAT) program designed by \citep{Wasser2004} and the BayEnv and BayEnv2 programs designed by \citep{Coop2010} and \citep{GuntherCoop2013}.  Under this model, the allele frequency of a local population deviates away from a global mean allele frequency specific to that locus, and populations covary, to varying extent, in their deviation from this global mean.  We model the strength of the covariance between two populations as a decreasing function of the geographic and ecological distance between them, so that populations that are closer in space or more similar in ecology tend to have more similar allele frequencies. 
We note that this model is not an explicit population genetics model, but a statistical model -- we fit the observed spatial pattern of genetic variation, rather than modeling the processes that generated it.
Informally, we can think of this model as representing the simplistic scenario of a set of spatially homogeneous populations 
at migration-drift equilibrium under isolation by distance.

The parameters of this model are estimated in a Bayesian framework using a Markov chain Monte Carlo algorithm \citep{Metropolis1953, Hastings1970}.  We demonstrate the utility of this method with two previously published datasets. The first is a dataset from several subspecies of \textit{Zea mays}, known collectively as teosinte \citep{Fang2012}, in which we examine the contribution of difference in elevation to genetic differentiation between populations. The second is a subset of the Human Genome Diversity Panel (HGDP, \citep{Conrad2006, Li2008}), for which we quantify the effect size of the Himalaya mountain range on genetic differentiation between human populations.  We have coded this method --- Bayesian Estimation of Differentiation in Alleles by Spatial Structure and Local Ecology (\emph{BEDASSLE}) --- in a user-friendly format in the statistical platform R \citep{R}, and have made the code available for download at \textit{genescape.org}.

\section*{Methods}

\subsection*{Data}

Our data consist of $L$ unlinked biallelic single nucleotide polymorphisms (SNPs) in $K$ populations; a matrix of pairwise geographic distance between the sampled populations ($D$); and one or more environmental distance matrices ($E$).  
The elements of our environmental distance matrix may be binary (e.g., same or opposite side of a hypothesized barrier to gene flow) or continuous (e.g., difference in elevation or average annual rainfall between two sampled populations).  
The matrices $D$ and $E$ can be arbitrary, so long as they are nonnegative definite, a constraint satisfied if they are each matrices of distances with respect to some metric.  
We summarize the genetic data as a set of allele counts ($C$) and sample sizes ($S$).  We use $C_{\ell,k}$ to denote the number of observations of one of the two alleles at biallelic locus $\ell$ in population $k$ out of a total sample size of $S_{\ell,k}$ alleles.  The designation of which allele is counted (for convenience, we denote the counted allele as  allele `1'), is arbitrary, but must be consistent among populations at the same locus.

\subsection*{Likelihood Function}

We model the data as follows.
The $C_{\ell,k}$ observed `1' alleles in population $k$ at locus $\ell$ result from randomly sampling a number $S_{\ell,k}$ of alleles from an underlying population in which allele 1 is at frequency $f_{\ell,k}$.  These population frequencies $f_{\ell, k}$ are themselves random variables, independent between loci 
but correlated between populations in a way that depends on pairwise geographic and ecological distance.  
A flexible way to model these correlations is to assume that the allele frequencies $f_{\ell,k}$ are multivariate normal random variables, inverse logit-transformed to lie between 0 and 1.  
In other words, we assume that $f_{\ell,k}$ 
is obtained by adding a deviation $\theta_{\ell,k}$ to the global value $\mu_\ell$,
and transforming:
\begin{equation} \label{eqn:logit_defn}
f_{\ell,k} =f(\theta_{\ell,k}+\mu_{\ell}) = \frac{1}{1+\exp(-(\theta_{\ell,k}+\mu_{\ell}))} .
\end{equation}
Under this notation, $\mu_{\ell}$ is the transformed mean allele frequency at locus $\ell$ and $\theta_{\ell,k}$ is the population- and locus-specific deviation from that transformed mean.
We can then write the binomial probability of seeing $C_{\ell,k}$ of allele '1' at locus $\ell$ in population $k$ as
\begin{equation}
  P\big(C_{\ell,k}|S_{\ell,k},f_{\ell,k} \big) = \binom{S_{\ell,k}}{C_{\ell,k}} f_{\ell,k}^{C_{\ell,k}}(1-f_{\ell,k})^{S_{\ell,k}-C_{\ell,k}}.
\end{equation}
In doing so, we are assuming that the individuals are outbred, so that the $S_{\ell,k}$ alleles represent independent draws from this population frequency.  We will return to relax this assumption later.

To model the covariance of the allele frequencies across populations, we assume that $\theta_{\ell,k}$ are multivariate normally distributed, with mean zero and a covariance matrix $\Omega$ that is a function of the pairwise geographic and ecological distances between the sampled populations.  We model the covariance between populations $i$ and $j$ as
\begin{equation} \label{eqn:covariance_form}
\Omega_{i,j} = \frac{1}{\alpha_{0}}\exp{\left(-(\alpha_{D}D_{i,j}+\alpha_{E}E_{i,j})^{\alpha_{2}}\right)} ,
\end{equation}
where $D_{i,j}$ and $E_{i,j}$ are the pairwise geographic and ecological distances between populations $i$ and $j$, respectively, and $\alpha_{D}$ and $\alpha_{E}$ are the effect sizes of geographic distance and ecological distance, respectively. The parameter $\alpha_{0}$ controls the variance of population specific deviate $\theta$ (i.e.\ at $D_{i,j} + E_{i,j} = 0$), and $\alpha_{2}$ controls the shape of the decay of the covariance with distance.   As alluded to above, as many separate ecological distance variables may be included as desired, each with its own $\alpha_{E_{x}}$ effect size parameter, but here we restrict discussion to a model with one.

With this model, 
writing $\alpha = (\alpha_{0},\alpha_{D},\alpha_{E},\alpha_{2})$,
the likelihood of the SNP counts observed at locus $\ell$ in all sampled populations can now be expressed as 
\begin{equation}
P \big( C_{\ell},\theta_{\ell}|S_{\ell},\mu_{\ell},\alpha \big) = P \big (\theta_{\ell}|\Omega(\alpha) \big) \prod_{k=1}^K P\big( C_{\ell,k}|S_{\ell,k},f(\theta_{\ell},\mu_{\ell}) \big )
\end{equation}
where 
we drop subscripts to indicate a vector (e.g.\ $C_\ell = (C_{\ell 1}, \ldots, C_{\ell K})$), 
and $P(\theta_{\ell}|\Omega)$ is the multivariate normal density with mean zero and covariance matrix~$\Omega$.

The joint likelihood of the SNP counts $C$ and the transformed population allele frequencies $\theta$ across all $L$ unlinked loci in the sampled populations is just the product across loci:
\begin{equation}
P \big( C,\theta|S,\mu,\alpha \big) =  \prod_{\ell=1}^L  P \big(\theta_{\ell}|\Omega(\alpha) \big) \prod_{k=1}^K P \big( C_{\ell,k}|S_{\ell,k},f(\theta_{\ell},\mu_{\ell}) \big) .
\end{equation}

\subsection*{Posterior Probability}
We take a Bayesian approach to inference on this problem, and specify priors on each of our parameters.  
We place exponential priors on $\alpha_D$ and $\alpha_E$, each with mean 1; and a gamma prior on $\alpha_0$, with shape and rate parameters both equal to 1.
We took the prior on $\alpha_2$ to be uniform between 0.1 and 2.
Finally, we chose a Gaussian prior for each $\mu_\ell$, with mean $0$, variance $1/\beta$, and a gamma distributed hyper-prior on $\beta$ with shape and rate both equal to 0.001.
For a discussion of the rationale for these priors, please see the Appendix.

The full expression for the joint posterior density, including all priors, is therefore given by 
\begin{align} \label{eqn:posterior_density}
  P(\theta,\mu,\alpha_{0},\alpha_{D},\alpha_{E},\alpha_{2},\beta|C, S) \propto \begin{split}
    \left( \prod_{\ell=1}^{L} P(\theta_{\ell,k}|\Omega)P(\mu_{\ell}|\beta )  \prod_{k=1}^K P(C_{\ell,k}|S_{\ell,k},f_{\ell,k}) \right) \\
 \qquad \qquad \times P(\beta)P(\alpha_{0})P(\alpha_{D})P(\alpha_{E})P(\alpha_{2})
 \end{split}
\end{align}
where the various $P$ denote the appropriate marginal densities, 
and the proportionality is up to the normalization constant given by the right-hand side integrated over all parameters.  

\subsection*{Markov chain Monte Carlo}
We wish to estimate the posterior distribution of our parameters, particularly $\alpha_{D}$ and $\alpha_{E}$ (or at least, their ratio).  As the integral of the posterior density given above cannot be solved analytically, we use Markov chain Monte Carlo (MCMC) to sample from the distribution.  We wrote a custom MCMC sampler in the statistical platform R \citep{R}.   
The details of our MCMC procedure are given in the Appendix.

\subsection*{Model Adequacy}
Our model is a simplification of the potentially complex relationships present in the data, and there are likely other correlates of differentiation not included in the model. Therefore, it is important to test the model's fit to the data, 
and to highlight features of the data that the model fails to capture.
To do this, we use posterior predictive sampling, using the set of pairwise population $F_{ST}$ values as a summary statistic \citep{WeirHill2002}, 
as we are primarily interested in the fit to the differentiation between pairs of populations.  In posterior predictive sampling, draws of parameters are taken from the posterior and used to simulate new datasets, summaries of which can be compared to those observed in the original datasets \citep{Gelman1996}.

Our posterior predictive sampling scheme proceeds as follows. For each replicate of the simulations we
\begin{enumerate}
\item Take a set of values of $\beta$ and all $\alpha$ parameters from their joint posterior (i.e.\ our MCMC output). 
\item Compute a covariance matrix $\Omega$ from this set of $\alpha$ and the pairwise geographic and ecological distance matrices from the observed data.
\item Use $\Omega$ to generate $L$ multivariate normally distributed $\theta$, and use $\beta$ to generate a set of normally distributed $\mu$.  
  These $\theta$ and $\mu$ are transformed using equation~\eqref{eqn:logit_defn} into allele frequencies for each population-locus combination, and binomially distributed allele counts are sampled using those frequencies and the per-population sample sizes from the observed data.
\item Calculate $F_{ST}$ between each pair of populations across all loci using the count data. Specifically we use the $F_{ST}$ estimator defined by the equation given on the top of page 730 in \citet{WeirHill2002}.
\end{enumerate}

We then use various visualizations of $F_{ST}(i,j)$, e.g.~plotted against distance between $i$ and $j$, to compare the patterns in the observed dataset to the patterns in the simulated datasets.  This functions as a powerful and informative visual summary of the ability of the model to describe the observed data.  Since $F_{ST}$ is a good measure of genetic differentiation, users can assess how well the method is able to pick up general trends in the data (e.g., increasing genetic differentiation with ecological or geographic distance) and how well those general trends in the model match the slope of their observed counterparts, and also identify specific pairwise population comparisons that the model is doing a poor job describing.  These latter may help reveal other important processes that are generating genetic differentiation between populations, such as unmeasured ecological variables, or heterogeneity in population demography.  

\subsection*{Accounting for overdispersion}

A consequence of the form of the covariance given in equation \eqref{eqn:covariance_form} is that all populations have the same variance of allele frequencies about the global mean
(and this is $\Omega_{ii}=1/\alpha_0$).
This will be the case in a homogeneous landscape,
but is not expected under many scenarios, such as those characterized by local differences in population size, inbreeding rate, historical bottlenecks, or population substructure.
In practice, this leads to overdispersion -- particular populations deviating more from global means than others.
Indeed, in both empirical datasets examined in this paper, there are clearly populations 
with much greater deviation in allele frequencies from the global mean
than predicted from their geographical and ecological distances.

To account for this, we will explicitly model the within-population correlations in allelic identity due to varying histories.
In so doing, we simultaneously keep outlier populations from having an undue influence on our estimates of $\alpha_D$ and $\alpha_E$, 
the effect sizes of the distance variables measured, and highlight those populations that the model is describing poorly.
Introducing correlations accounts for overdispersion because a population whose allele frequencies differ more from its predicted frequencies across loci
has individuals whose allelic identities are more correlated (and the converse is also true).
To see this, observe that, for instance, if one completely selfing population and one outbred population each have a given allele at frequency $p$, then the variance in sampled allele frequency will be twice as high in the selfing population,
since the number of effective independent draws from the pool of alleles is half as large.

To introduce within-population correlations 
we assume that the allele frequencies from which the allele counts $C_{\ell,k}$ are drawn are not fixed at $f_{\ell,k}$,
but rather randomly distributed, with mean given by $f_{\ell,k}$ and variance controlled by another parameter.
Specifically, given $\mu_\ell$ and $\theta_{\ell,k}$, we suppose that the allele frequency at locus $\ell$ in population $k$ is beta-distributed 
with parameters $\Phi_k f_{\ell,k}$ and $\Phi_k (1-f_{\ell,k})$, where $f_{\ell,k}=f(\mu_\ell,\theta_{\ell,k})$ as before, and $\Phi_k$ is a population-specific parameter, estimated separately in each population, that controls the extent of allelic correlations between draws from individuals in population $k$.
To see why this introduces allelic correlations, consider the following equivalent description of the distribution of $C_{\ell,k}$.
We sample the alleles one at a time; 
if we have drawn $n$ alleles; then the $(n+1)^\mathrm{st}$ allele is either:
a new draw with probability $\Phi_k/(\Phi_k+n)$ (in which case it is of type `1' with probability $f_{\ell,k}$ and of type '0' with probability $1-f_{\ell,k}$);
otherwise, it is of the same type as a previously sampled allele, randomly chosen from the $n$ sampled so far.
Conceptually,
each allele is either a ``close relative'' of an allele already sampled, or else a ``new draw'' from the ``ancestral population'' with allele frequency $f_{\ell,k}$.
Smaller values of $\Phi_k$ lead to increased allelic correlations, which in turn increase the variance of population allele frequencies.

Conveniently, the random frequency integrates out, so that the likelihood of the count data becomes
\begin{equation}
  P(C_{\ell,k}|S_{\ell,k},f_{\ell,k} =f(\theta_{\ell,k},\mu_{\ell})) = \binom{S_{\ell,k}}{ C_{\ell,k}} \frac{ B( C_{\ell,k} + \Phi_{k}f_{\ell,k} , S_{\ell,k}-C_{\ell,k} + \Phi_{k}(1-f_{\ell,k}) )}{ B ( \Phi_{k}f_{\ell,k}  , \Phi_{k}(1-f_{\ell,k}) )},
\end{equation}
where $B(x,y)$ is the beta function. 
This is known as the ``beta-binomial'' model \citep{williams1975analysis},
and is used in a population genetics context by \citet{BaldingNichols1995,BaldingNichols1997}; see \citet{Balding2003} for a review.

The parameter $\Phi_k$ can be related to one of Wright's $F$-statistics \citep{Wright1943}.  
As derived in previous work \citep{BaldingNichols1995,BaldingNichols1997}, 
if we define $F_k$ by $\Phi_k=F_{k}/(1-F_{k})$ ($0 \leq F_k <1$), 
then $F_{k}$ is analogous to the inbreeding coefficient for population $k$ relative to its set of the spatially predicted population frequencies \citep{Cockerham1986, Balding2003},
with higher $F_{k}$ corresponding to higher allelic correlation in population $k$, as one would expect given increased drift (inbreeding) in that population.
However, it is important to note that $F_{k}$ cannot solely be taken as an estimate of the past strength of drift,
since higher $F_{k}$ would also be expected in populations that simply fit the model less well.
We report values of $F_k$ in the output and results, and discuss the interpretation of this parameter further in the discussion.

We have coded this beta-binomial approach as an alternative to the basic model (see Results for a comparison of both approaches on empirical data).  To combine estimation of this overdispersion model into our inference framework, we place an inverse exponential prior on $\Phi_k$ (that is, $1/\Phi_k \sim \Exp(5)$). This prior and the beta-binomial probability density function are incorporated into the posterior.  

\subsection*{Simulation Study}

We conducted two simulation studies to evaluate the performance of the method.  In the first, we simulated data under the inference model, and in the second, we simulated under a spatially explicit coalescent model.

\begin{figure}[ht!]
\begin{center}
  \includegraphics[width=6in,height=5.14in]{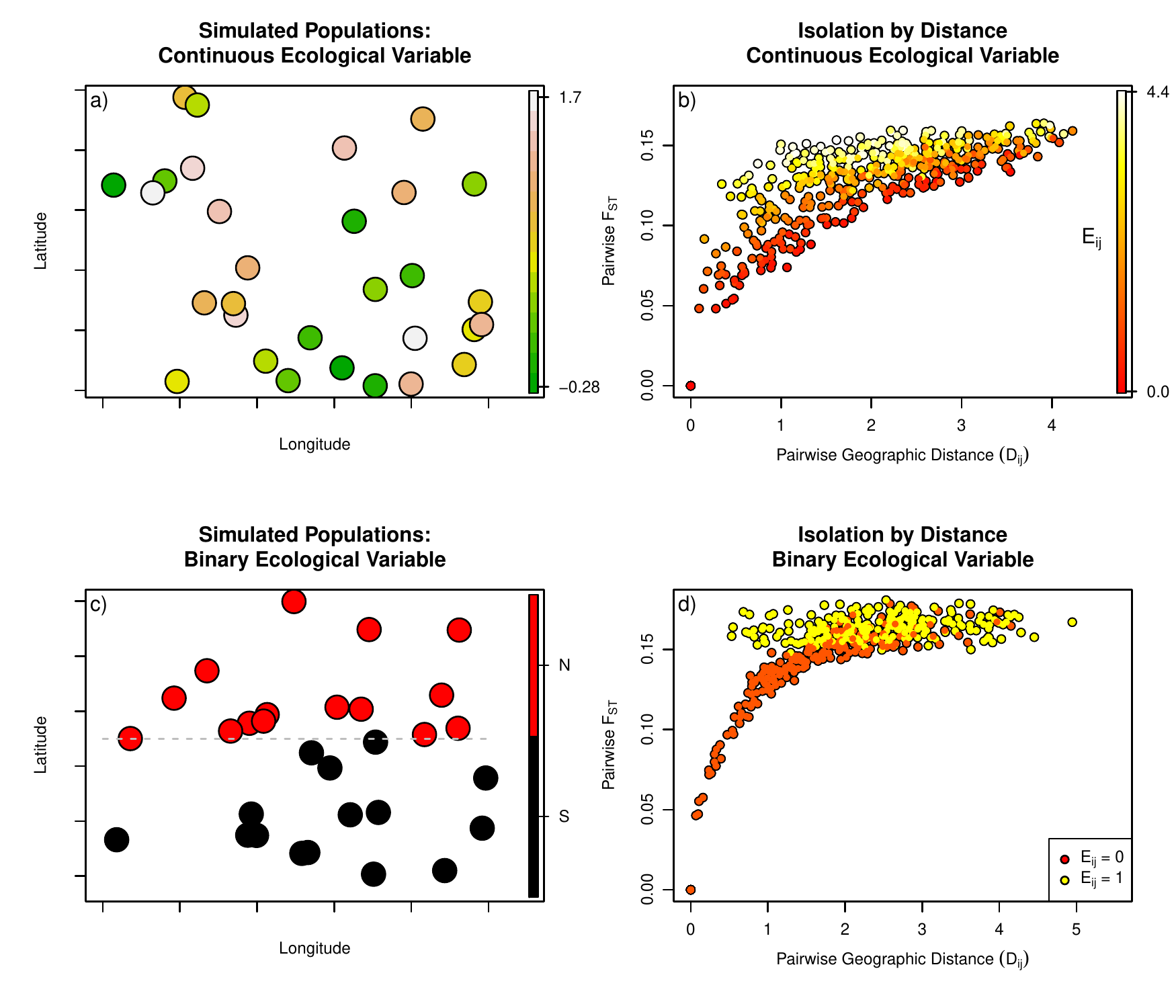}
 \caption{
	\bf{a)}
 		\textmd{Populations simulated in the unit square, colored by their value of a continuous ecological variable.}
	\bf{b)}
 		\textmd{Pairwise $F_{ST}$ between simulated populations from (a), colored by difference in their values of the continuous ecological variable.}
	\bf{c)}
  		\textmd{Populations simulated in the unit square, colored by their value of a binary ecological variable.}
	\bf{d)}
 		\textmd{Pairwise $F_{ST}$ between simulated populations from (c), colored by difference in their values of the binary ecological variable.}
 \label{sfig:sim_info}
  }
\end{center}
\end{figure}

For the datasets simulated under the model, 
each simulated dataset consisted of 30 populations, each with 10 diploid individuals sequenced at 1000 polymorphic bi-allelic loci.  Separately for each dataset, the geographic locations of the populations were sampled uniformly from the unit square, and geographic distances ($D_{i,j}$) were calculated as the Euclidean distance between them.  We also simulated geographically autocorrelated environmental variables, some continuous, some discrete (see Figure \ref{sfig:sim_info}\textit{a} and \textit{c}).  For both discrete and continuous variables we simulated datasets in which ecological distance had no effect on genetic differentiation between populations; these simulations tested whether our method avoids the false positive issues of the partial Mantel test.  We also simulated datasets with an effect of both geographic and ecological distance on genetic distance across a range of relative effect sizes (varying the ratio $\alpha_{E}/\alpha_{D}$) to test our power to detect their relative effects. The study thus consisted of four sections, each comprised of 50 datasets: discrete and continuous ecological variables, with or without an effect of ecology.

For each dataset, we set $\alpha_{0} =0.5$, and sampled $\alpha_{D}$ and $\alpha_2$ from uniform distributions ($U(0.2,4)$ and $U(0.1,2)$ respectively); the choice of $\alpha_E$ varied, depending on the specific scenario (described below).  These parameters were chosen to give a range of pairwise population $F_{ST}$ spanning an order of magnitude between approximately 0.02 and 0.2, and a realistic allele frequency spectrum.
The covariance matrix $\Omega$ was calculated using these $\alpha$ and the pairwise geographic and ecological distance matrices (normalized by their standard deviations), 
and $\Omega$ was used to generate the multivariate, normally distributed $\theta$. 
Values of $\mu$ were drawn from a normal distribution with variance $1/(\beta=0.09)$.
Allele frequencies at each locus were calculated for each population from the $\theta$ and $\mu$ using equation~\eqref{eqn:logit_defn}, and SNP counts at each locus in each population were drawn from binomial distributions parameterized by that allele frequency with the requirement that all loci be polymorphic.   We simulated under the following ecological scenarios.

\paragraph{1. Continuous, Autocorrelated Ecological Variable}
For the continuous case, we simulated the values of an ecological variable across populations by sampling from a multivariate normal distribution with mean zero and covariance between population $i$ and population $j$ equal to $\cov(E(i),E(j)) = \exp(-D_{i,j}/a_{c})$, where $a_{c}$ determines the scale of the autocorrelation \citep[following][]{Guillot2013}.  For all simulations, we set $a_{c}= 0.7$, 
to represent a reasonably distributed ecological variable on a landscape.

\paragraph{2. Binary Ecological Variable} A binary variable was produced by declaring that the latitudinal equator in the unit square was a barrier to dispersal, so that all populations on the same side of the barrier were separated by an ecological distance of zero, and all population pairs that spanned the equator were separated by an ecological distance of 1.

\paragraph{A. Zero Effect Size} For each type of ecological variable, we produced 50 simulated datasets with $\alpha_{E}=0$, so that ecological distance had no effect on the covariance of $\theta$, and hence on genetic differentiation between populations.  For each of these simulated datasets, we performed a partial Mantel test in R using the package \textit{ecodist} \citep{ecodist} with 1,000,000 permutations. 
	
\paragraph{B. Varying Effect Size} We also produced 50 simulated datasets for each type of ecological variable by simulating ten datasets for each value of $\alpha_{E}/\alpha_{D}$ from 0.2 to 1.0 in intervals of 0.2 (see Figure \ref{sfig:sim_info}\textit{b} and \textit{d}).  (As above, values of $\alpha_{D}$ were drawn from a uniform distribution ($U(0.2,4)$), so this determines $\alpha_E$.) 

\begin{figure}[ht!]
\begin{center}
  \includegraphics[width=3in,height=3.5in]{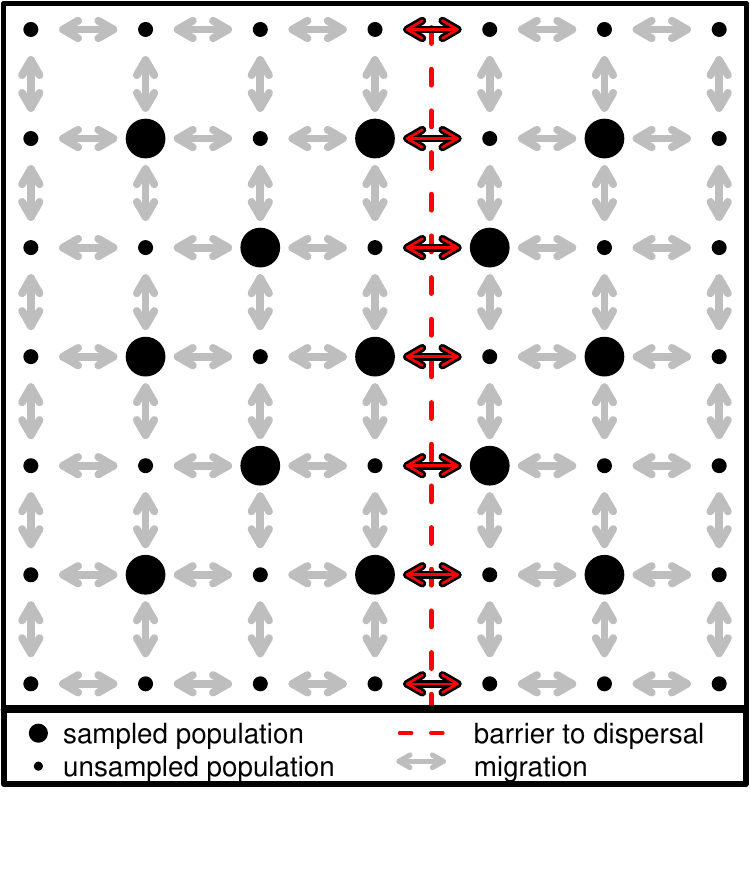}
 \caption{
 		\textmd{Populations simulated using a spatially explicit coalescent model in the unit square.  All simulated populations are indicated with black dots, while populations that were sampled for inclusion in each dataset are indicated by large black dots.  All pairwise migration is indicated with gray arrows.  The barrier to dispersal is given by the red dotted line, across which the standard migration rate was divided by a barrier effect size, which we varied.}
	 \label{fig:ms_lattice}
  }
\end{center}
\end{figure}

For the datasets simulated using a spatially explicit coalescent process, allelic count data were simulated on a fixed lattice using the program \textit{ms} \citep{Hudson2002}.  A total of 49 populations were simulated, evenly spaced in a seven-by-seven grid, of which a subset of 25 populations were sampled to make the final dataset; these 25 sampled populations were arranged in a five-by-five grid, as shown in Figure \ref{fig:ms_lattice}.  Each population consisted of 10 chromosomes sampled at 1,000 polymorphic, unlinked, biallelic loci.  Migration occurred between neighboring populations (with no diagonal migration) at a rate of $4Nm_{i,j} = 4$.  In all simulations, a longitudinal potential barrier to gene flow was included just to the east of the central line (see Figure \ref{fig:ms_lattice}).  Migration rate between populations that were separated by this barrier was diminished by dividing by some barrier effect size, which varied between simulation sets.  For 40 datasets, the barrier effect size was set to 1, so that the barrier had no effect on genetic differentiation across it.  The barrier effect size was set to 5, 10, and 15, for 20 datasets each, for a total of 100 datasets simulated under the spatial coalescent.  For all datasets, geographic distance was measured as the pairwise Euclidean distance between populations on the lattice, and ecological distance was defined as zero between populations on the same side of the barrier, and 1 between populations on opposite sides.

\paragraph{}
All analyses on the simulated datasets were run for 1,000,000 MCMC iterations, which appeared sufficient in most cases for convergence on the stationary distribution.  The chain was sampled every 1,000 generations, and all summary statistics from the simulation study were calculated after a burn-in of 20\%.  The metrics of method performance used on the datasets simulated under the inference model were precision, accuracy, and coverage of the $\alpha_{E}:\alpha_{D}$ ratio.  We defined \emph{precision} as breadth of the 95\% credible set of the marginal posterior distribution; \emph{accuracy} as the absolute value of the difference between the median value of the marginal posterior distributions and the values used to simulate the data in each dataset; and \emph{coverage} as the proportion of analyses for which the value used to simulate the data fell within the 95\% credible set of the marginal posterior distribution for that parameter.  For the datasets simulated under the spatial coalescent process, we wished to assess the ability of the method to accurately recover the relative strength of the barrier to gene flow.

For approximately 30\% of all analyses, the MCMC runs displayed obvious difficulty with convergence within the first 1,000,000 generations.  The signs of potentially poor single-chain MCMC behavior that we looked for included: acceptance rates that are too low or too high (generally 20-70\% acceptance rates are thought to be optimal); parameter trace plots that exhibit high autocorrelation times; acceptance rates that have not plateaued by the end of the analysis; and marginal distributions that are multimodal, or not approximately normal (for a more complete discussion on MCMC diagnosis, please see \citet{Gilks1996}; for plots of example MCMC output, see Figures \ref{sfig:trace_plots}, \ref{sfig:joint_marginals}, and \ref{sfig:acceptance_rates}).  In some cases, this was because the naive scales of the various tuning parameters of the random-walk proposal mechanisms were inappropriate for the particular dataset, and mixing was too slow over the number of generations initially specified (as diagnosed by visualizing the parameter acceptance rates of MCMC generations).  This was addressed by re-running analyses on those datasets using different random-walk tuning parameters, or by increasing the number of generations over which the MCMC ran.  In the other cases, failure to converge was due to poor performance of the MCMC in regions of parameter space too near the prior boundaries.  Specifically, when the chain was randomly started at values of some $\alpha$ parameters too close to zero, it was unable to mix out of that region of parameter space.  This problem was addressed by re-running the analyses using different, randomly chosen initial values for the $\alpha$ parameters.  In our R package release of the code we provide simple diagnostic tools for the MCMC output, and further guidance for their use.

\subsection*{Empirical Data}
To demonstrate the utility of this method, we applied it to two empirical datasets: one consisting of populations of teosinte (\textit{Zea mays}), the wild progenitor of maize, and one consisting of human populations from the HGDP panel.  Both processed datasets are available for download at \emph{genescape.org}. 
See Tables S1 and S2 in the Supplementary Materials  for names and metadata of populations used. 

The teosinte dataset consisted of 63 populations of between 2 and 30 diploid individuals genotyped at 978 biallelic, variant SNP loci \citep{Fang2012}.  Each population was associated with a latitude, longitude, and elevation at the point of sampling (see Figure \ref{sfig:zea_Fk_map} and Table S1).  Pairwise geographic great-circle distances and ecological distances were calculated for all pairs of populations, where ecological distance was defined as the difference in elevation between populations.  Both pairwise distance variables were normalized by their standard deviations.  

The human dataset was the Eurasian subset of that available from the HGDP \citep{Conrad2006, Li2008}, consisting of 33 populations of between 6 and 45 individuals genotyped at 1000 biallelic, variant SNP loci (see Figure \ref{sfig:him_Fk_map} and Table S2).  Pairwise geographic great-circle distances and ecological distances were calculated for all pairs of populations, where ecological distance was defined as 0 or 1 if the populations were on the same or opposite side of the Himalaya mountain range, respectively. For the purposes of our analysis the western edge of the Himalaya was defined at $75^{\circ}$ East. 

For comparison, the method was run on each of the two datasets both with and without the beta-binomial overdispersion model.  MCMC marginal traces were examined visually to assess convergence on a stationary distribution.  The chain was thinned by sampling every 1000 generations, and the median and 95\% credible sets were reported on the marginal distribution after a burn-in of 20\%.  The MCMC analysis for the teosinte dataset without the overdispersion model was run for 10 million generations; the analysis with the overdispersion model was run for 15 million generations.  For the HGDP dataset, the numbers of generations were 25 million and 35 million, for the analyses without and with the overdispersion model, respectively.

\section*{Results}
\subsection*{Simulation Results}

As described above, we conducted two simulation studies.  The performance of the method in inference of the parameters of greatest interest is given below.   

First we note that,
consistent with the results of \citep{Guillot2013}, the spatial autocorrelation in our ecological variable caused the partial Mantel to have a high false positive rate when $\alpha_E=0$, which suggests that the partial Mantel test is not well calibrated to assess the significance of ecological distance on patterns of genetic differentiation.  At a significance level of $p=0.05$, the false positive rate for the datasets simulated under the inference model with a binary ecological distance variable was 8\%, and for the continuous ecological variable, the false positive error rate was 24\%.  For the datasets simulated under the spatial coalescent process with a barrier effect size of 1 (meaning that the barrier had no effect on genetic differentiation across it), the false positive error rate was 37.5\% (see Figure \ref{sfig:Pmantel}).

\begin{figure}[ht!]
\begin{center}
  \includegraphics[width=6in,height=3.42in]{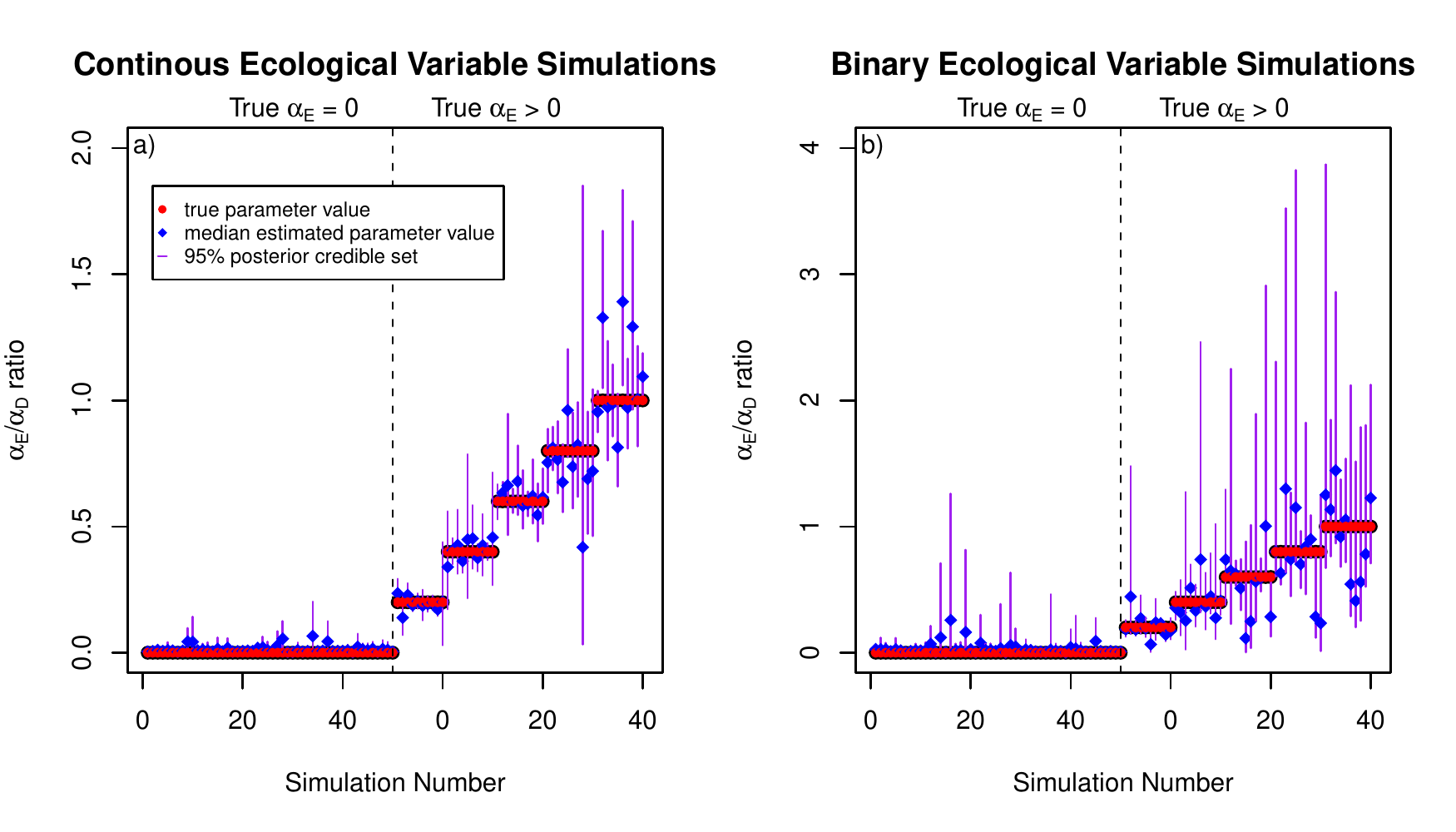}
 \caption{
	\textbf{a)}
 		Performance of the method for the 100 datasets simulated with a continuous ecological distance variable.  
	\textbf{b)}
 		Performance of the method for the 100 datasets simulated with a binary ecological distance variable.  
		In each, the left panel depicts performance on the 50 datasets for which $a_{E}$ was fixed at 0, and the right panel depicts performance on the 50 datasets for which $a_{E}$ varied.
 \label{sfig:sim_summs}
  }
\end{center}
\end{figure}

The precision and accuracy results for the datasets simulated under the model with a continuous and discrete ecological variable are visualized in Figure panels \ref{sfig:sim_summs}\textit{a} and \textit{b}, respectively, across the six simulated values of the ratio $\alpha_E/\alpha_D$.
Median precision, accuracy, and coverage are reported in Table~\ref{tab:sim_summs}.  

\begin{figure}[ht!]
\begin{center}
  \includegraphics[width=6in,height=4in]{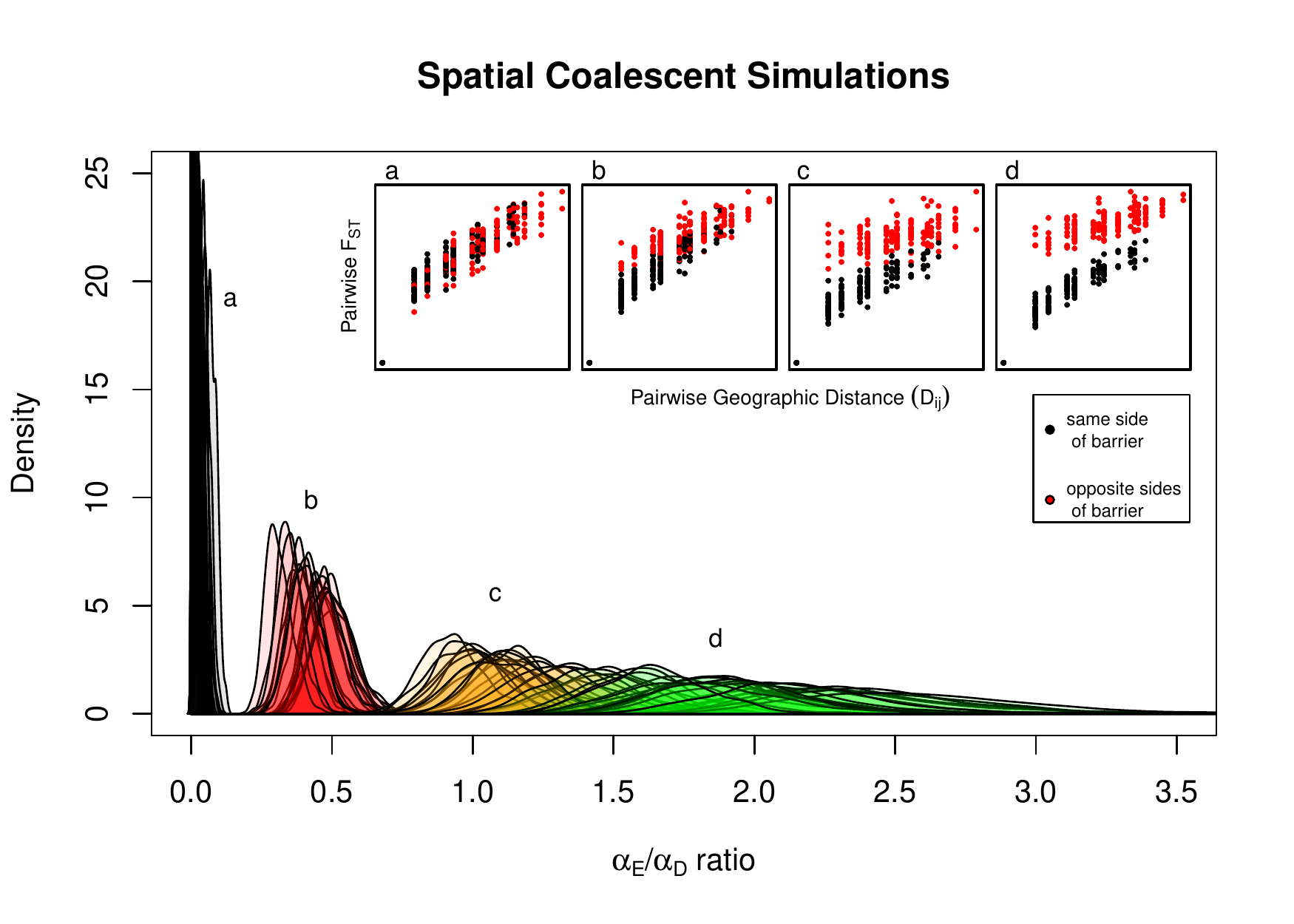}
 \caption{
 		The marginal distributions on the $\alpha_E/\alpha_D$ ratio from the analyses performed on the datasets simulated using a spatially explicit coalescent process.  The migration rate between populations separated by the barrier was divided by a barrier effect size, which varied among simulations.  \bf{Inset:}  \textmd{Pairwise $F_{ST}$, colored by whether populations were on the same or opposite sides of a barrier to dispersal, plotted against pairwise geographic distance for example datasets for each of the 4 barrier effect sizes.}
	\bf{a)}
 		\textmd{Barrier effect size of 1 (n=40);}
	\bf{b)}
 		\textmd{Barrier effect size of 5 (n=20);}
	\bf{c)}
 		\textmd{Barrier effect size of 10 (n=20);}
	\bf{d)}
 		\textmd{Barrier effect size of 15 (n=20).}
 \label{sfig:fig5_allmarg}
 }
\end{center}
\end{figure}

The performance of the method on the datasets simulated using the spatial coalescent model is given in Figure \ref{sfig:fig5_allmarg}, which shows the posterior distributions of $\alpha_E:\alpha_D$ ratio from each analyzed dataset over the four barrier effect sizes.

\begin{table}[ht!!]
\begin{center}
\begin{tabular}{| l || c | c | c | c |}
\hline
& Sim Study 1A & Sim Study 1B & Sim Study 2A & Sim Study 2B \\ \hline
Precision & 0.041 & 0.30 & 0.15 & 0.96 \\ \hline
Accuracy & 0.013 & 0.0066 & 0.031 & 0.033 \\ \hline
Coverage & NA & 94\% & NA & 94\% \\
\hline
\end{tabular}  
\end{center}
\caption{
Simulation Studies 1A and 1B were conducted with a continuous ecological variable and $\alpha_E = 0$ and $\alpha_E > 0$, respectively.  Simulation Studies 2A and 2B were conducted with a binary ecological variable and $\alpha_E = 0$ and $\alpha_E > 0$, respectively.  Precision, accuracy, and coverage are reported on inference of the $\alpha_E$ : $\alpha_D$ ratio. \emph{Precision} is breadth of the 95\% credible set of the marginal posterior distribution (smaller values indicate better method performance).  \emph{Accuracy} is the absolute value of the difference between the median value of the marginal posterior distributions and the values used to simulate the data (smaller values indicate better method performance).  
\emph{Coverage} is the proportion  of analyses for which the value used to simulate the  data fell within the 95\% credible set of the marginal posterior distribution for that parameter (higher values indicate better method performance).  Coverage is not reported for the simulations in which the effect size of the ecological distance variable was fixed to zero ($\alpha_E = 0$), as the parameter value used to generate the data is on the prior bound on $\alpha_{E}$, and coverage was therefore zero.}
 \label{tab:sim_summs}
\end{table}

\subsection*{Empirical Results}
\subsection*{Teosinte Results}
For the \emph{Zea mays} SNP dataset analysis, the mean and median of the posterior ratio of the effect size of pairwise difference in elevation to the effect size of pairwise geographic distance (i.e.- the $\alpha_{E} : \alpha_{D}$ ratio) was 0.153, and the 95\% credible set was 0.137 to 0.171 (see Figure ~\ref{sfig:zea_traceplot}\textit{a}).  The interpretation of this ratio is that one thousand meters of elevation difference between two populations has a similar impact on genetic differentiation as around 150 (137--171) kilometers of lateral distance. 

Accounting for overdispersion (using the beta-binomial model) we obtain slightly different results, with a mean and median $\alpha_{E} : \alpha_{D}$ ratio of 0.205, and a 95\% credible set from 0.180 to 0.233 (1,000 meters difference in elevation $\approx$ 205 kilometers lateral distance, see Figure ~\ref{sfig:zea_traceplot}\textit{b}).  
Values of our $F$ statistics $F_{k}$ estimated across populations ranged from $2\times 10^{-4}$ to 0.53, and are shown in Supplemental Figure \ref{sfig:zea_Fk_map}.  

\begin{figure}[ht!]
\begin{center}
  \includegraphics[width=6in,height=5.14in]{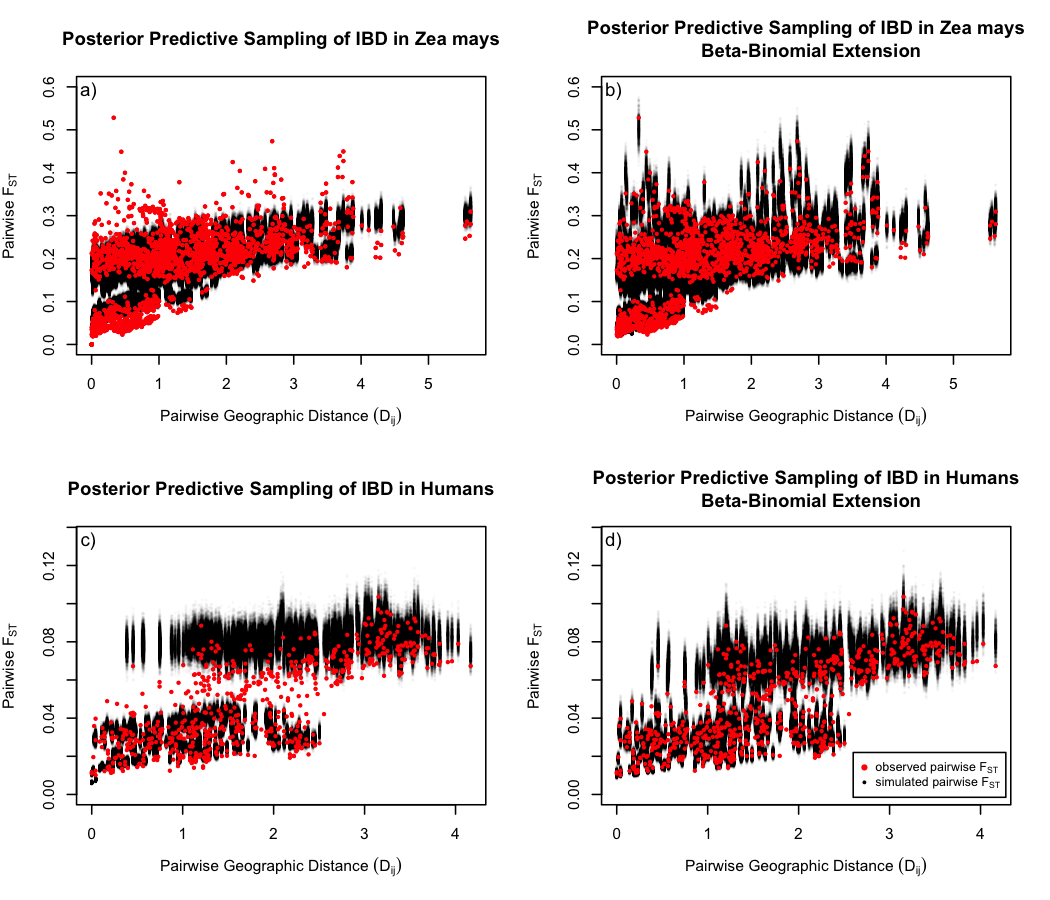}
 \caption{
 		\textmd{Posterior predictive sampling with 1,000 simulated datasets, using pairwise $F_{ST}$ as a summary statistic of the allelic count data for:}
	\bf{a)}
 		\textmd{the teosinte dataset, using the standard model;}
	\bf{b)}
 		\textmd{the teosinte dataset, using the overdispersion model;}
	\bf{c)}
 		\textmd{HGDP dataset, standard model.}
	\bf{d)}
 		\textmd{HGDP dataset, overdispersion model.}
 \label{sfig:pps}
  }
\end{center}
\end{figure}

Posterior predictive sampling indicates incorporating overdispersion with the beta-binomial extension results in a better fit to the data (see Figure \ref{sfig:pps}\textit{a} and \textit{b}): the mean Pearson's product moment correlation between the posterior predictive datasets and the observed data without the beta-binomial extension was 0.64, while the mean correlation with the beta-binomial model was 0.86 (see Figure \ref{sfig:pps.corr}\textit{a}).  The ability of the model to predict specific pairwise population $F_{ST}$ is shown Figure \ref{sfig:zea.pps.pval}.  

\subsection*{HGDP results}  
For the human (HGDP) SNP dataset analysis, the mean posterior $\alpha_{E} : \alpha_{D}$ ratio was $5.13 \times 10^{4}$, the median was $5.00 \times10^{4}$, and the 95\% credible set was $3.09 \times10^{4}$ to $7.85\times10^{4}$ (see Figure \ref{sfig:him_traceplot}\textit{a}). However, this result seems to be sensitive to outlier populations, as the beta-binomial extension of this method on the same dataset yields significantly different results, with a mean $\alpha_{E} : \alpha_{D}$ ratio of $1.35\times10^{4}$, a median of $1.34\times10^{4}$, and a 95\% credible set from $1.09\times10^{4}$ to $1.65\times10^{4}$ (see Figure \ref{sfig:him_traceplot}\textit{b}). 
This latter result is broadly consistent with that of \citet{Rosenberg2011}, who found an effect size ratio of $9.52\times10^{3}$ in a linear regression analysis that treated pairwise population comparisons as independent observations. The interpretation of our result is that being on the opposite side of the Himalaya mountain range has the impact of between approximately 11 and 16 thousand kilometers of extra pairwise geographic distance on  genetic differentiation. 

Under our beta-binomial extension values of $F_{k}$ estimated across populations ranged from $3.2 \times10^{-4}$ to 0.06.  Population values of $F_{k}$ are shown on the map in Figure \ref{sfig:him_Fk_map}. 

Posterior predictive sampling again indicates a better fit to the data including overdispersion (see Figure \ref{sfig:pps}\textit{c} and \textit{d}): the mean Pearson's product moment correlation between the posterior predictive datasets and the observed data without the beta-binomial extension was 0.88, while the mean correlation with the beta-binomial model was 0.91 (see Figure \ref{sfig:pps.corr}\textit{b}).  The ability of the model to predict specific pairwise population $F_{ST}$ is shown in Figure \ref{sfig:him.pps.pval}.  

\section*{Discussion}

In this paper, we have presented a method that uses raw allelic count data to infer the relative contribution of geographic and ecological distance to genetic differentiation between sampled populations. 
The method performs quite well: we have shown that it reliably and accurately estimates correct parameter values using simulations, and produces sensible models that give a good fit to observed patterns of differentiation in real datasets.
We feel that our method has broad utility to the field of landscape genetics and to studies of local adaptation, and holds a number of advantages over existing methods.  
(although see \citet{Wang2012} for another recent approach.)
It allows users to simultaneously quantify effect sizes of geographic distance and ecological distance (rather than assessing the significance of a correlation once the effect of geography has been removed, as in the partial Mantel test).  Explicitly modeling the covariance in allele frequencies allows users to accommodate non-independence in the data, and the method's Bayesian framework naturally accommodates uncertainty and provides a means of evaluating model adequacy.  
The inclusion of overdispersion allows fit to a set of populations with heterogeneous demographic histories.
In addition, the basic model presented here -- a parametric model of spatial covariance in allele frequencies -- is extremely versatile, allowing for the inclusion of multiple ecological or geographic distance variables, as well as great flexibility in the function used to model the covariance.

\subsection*{Simulation Study}

Our method performed well in both simulation studies (see Figure \ref{sfig:sim_summs}, Table \ref{tab:sim_summs}, and Figure \ref{sfig:fig5_allmarg}), and was able to effectively recognize and indicate when an ecological variable contributes significantly to genetic differentiation.  This is in contrast to the partial Mantel, which has a high false positive rate in the presence of spatial autocorrelation of environmental variables (see Figure \ref{sfig:Pmantel}).

For datasets simulated under the inference model, coverage, accuracy, and precision were all satisfactory (see Table \ref{tab:sim_summs}). The precision of our estimator of $\alpha_E$ was generally lower for our discrete ecological variable, likely due to the strong spatial structure of the discrete ecological variable.  

For datasets simulated using the spatial coalescent, there were no true values for the $\alpha_{E}:\alpha_{D}$ ratio to compare with those inferred by the method.  However, we note that the $\alpha_{E}:\alpha_{D}$ ratios estimated across analyzed simulated datasets tracked the barrier effect sizes used to simulate them, and that when the barrier had no effect on migration, the marginal distributions on the $\alpha_{E}:\alpha_{D}$ ratio estimated were stacked up against the prior bound at zero and had very low median values.  The width of the 95\% credible set of the marginal posteriors grew with the barrier effect size as a result of the flattening of the posterior probability surface as true parameter value increased.  Overall, the method performed well on the datasets simulated under a model different from that used for inference (and presumably closer to reality).

An issue we observed in practice is that at some parameter values, different combinations of $\alpha$ are essentially nonidentifiable --- the form of the covariance given in equation \eqref{eqn:covariance_form} sometimes allows equally reasonable fits at different values of $\alpha_2$, or at different combinations of $\alpha_0$, $\alpha_{D}$, and $\alpha_{E}$.  
(In other cases, all four parameters can be well-estimated.)
Even when this is the case, the $\alpha_E$ : $\alpha_D$ ratio, which is the real parameter of interest,
remains constant across the credible region, even as $\alpha_E$ and $\alpha_D$ change together to compensate for changes in $\alpha_2$ and $\alpha_0$.
Such `ridges' in the likelihood surface are readily diagnosed by viewing the trace plots and joint marginals of the $\alpha$ parameters (see Figures \ref{sfig:trace_plots} and \ref{sfig:joint_marginals}).

\subsection*{Empirical Results}

\subsubsection*{Teosinte}

The application of our method to the teosinte SNP dataset indicated that difference in elevation has a potentially substantial contribution to genetic differentiation between teosinte populations.  Difference in elevation could be correlated with another, as yet unmeasured ecological variable, so we cannot claim to report a causal link, but these results are certainly suggestive, especially in the light of the research on morphological adaptations in teosinte to high altitude \citep{EaglesLothrop1994}.  

The analysis of the teosinte SNP data with the beta-binomial extension of our method shows a much better model fit, and highlights a number of populations with particularly high $F_{k}$ values.  These populations (highlighted in Figure \ref{sfig:zea_Fk_map}) all belong to the subspecies \textit{Zea mays mexicana}, which primarily occurs at higher altitudes and is hypothesized to have undergone significant drift due to small effective population sizes or bottlenecks \citep{Fukunaga2005}.  In addition, a number of these populations occur in putative hybrid zones between \textit{Zea mays mexicana} and \textit{Zea mays parviglumis}, a separate, co-occuring subspecies \citep{vanHeerwaarden2011}.  Like drift, admixture would have the effect of increasing the variance in observed allele frequencies around the expectation derived from the strict geographic/ecological distance model, and would drive up the inferred $F_{k}$ parameters for admixed populations.  

\subsubsection*{HGDP}

In the Human Genome Diversity Panel data we find a strong effect of separation by the Himalayas on genetic differentiation, confirming previous results \citep[e.g.][]{Rosenberg2005}.
To obtain a good fit to the data it is necessary to model overdispersion (with the beta-binomial extension).  This lack of model fit of the basic model can be seen in the posterior predictive sampling in Figure \ref{sfig:pps}\textit{c} and \textit{d}, which highlights the importance of assessing model adequacy during analysis. 
Under the beta-binomial extension the $\alpha_E/\alpha_D$ ratio estimates an effect of the Himalayas far greater than the distance simply to circumnavigate around the Himalayas.  We think this likely reflects the fact that Eurasian populations are away from migration-selection equilibrium, reflecting past large-scale population expansions \citep{Keinan2007}.

 With overdispersion included, the model appears to describe the data reasonably well, suggesting substantial heterogeneity beyond that dictated by geographic distance and separation by the Himalayas between the sampled populations.  A number of populations stand out in their $F_{k}$ values, in particular the Kalash, the Lahu, the Mozabites, the Hazara, and the Uygur (highlighted in Figure \ref{sfig:him_Fk_map}). This is consistent with the known history of these populations and previous work on these samples \citep{Rosenberg2002}, which suggests that these populations are unusual for their geographic position (that is, they depart from expectations of their covariance in allele frequencies with their neighbors). The Hazara and Uygur populations are known to be recently admixed populations between central Asian and East ancestry populations.  The Mozabite population has substantial recent admixture from Sub-Saharan African populations \citep{Rosenberg2002,Rosenberg2011}.  The Kalash, who live in northwest Pakistan, are an isolated population with low heterozygosity, suggesting a historically small effective population size.  Finally the Lahu have unusually low heterozygosity compared to the other East Asian populations, suggesting that they too may have had an unusually low effective population size.  Thus our beta-binomial model, in addition to improving the fit to the data, is successfully highlighting populations that are outliers from simple patterns of isolation by distance.

\subsubsection*{Population-specific variance}
As noted above, in both empirical datasets analyzed, the beta-binomial extension to the basic model offers substantially better model fit. This could in part reflect ecological variables not included in the analyses, in addition to heterogeneity in demographic processes, both of which could shape genetic variation in these populations by pushing population allele frequencies away from their expectations under our simple isolation by distance and ecology model.  Our $F_{k}$ statistic provides a useful way to highlight populations that show the strongest deviations away from our model, and to prevent these deviations from obscuring environmental correlations or causing spurious correlations.  Therefore, we recommend that the extended model be used as the default model for analyses. 

\subsection*{Limitations}
The flexibility of this statistical model is accompanied by computational expense.  Depending on the number of loci and populations in a dataset, as well as the number of MCMC generation required to accurately describe the stationary distribution, analyses can take anywhere from hours to days.  Speedups could be obtained by parallelization or porting code to C.  In addition, as with any method that employs an MCMC algorithm, users should take care to assess MCMC performance to ensure that the chain is mixing well, has been run for a sufficient number of generations, and has converged on a stationary distribution \citep{Gilks1996}.  Users are well advised to run multiple independent chains from random initial locations in parameter space, and to compare the output of those analyses to confirm that all are describing the same stationary distributions. 

Our model rests on a number of assumptions, principal among which is that population allele frequencies are well-represented by a spatially homogeneous process, such as are obtained under mutation-migration equilibrium.  That is, we assume that current patterns of gene flow between populations are solely responsible for observed patterns of genetic differentiation.  
Some examples of biological situations that may violate the assumptions of our model include: two populations that have higher genetic differentiation than expected based on their pairwise geographic distance because they arrived in nearby locations as part of separate waves of colonization;  or two populations that have been recently founded on either side of some landscape element that truly does act as a barrier to gene flow, but that do not exhibit strong genetic differentiation yet, because the system is not in equilibrium.  In reality, we expect that very few natural populations will conform perfectly to the assumptions of our model; however, we feel that the method will provide valid approximations of the patterns for many systems, and that it will be a useful tool for teasing apart patterns of genetic variation in populations across heterogeneous landscapes.

\subsection*{Extensions}

The flexibility of this method translates well into extendability.  Among a number of natural extensions the community might be interested in implementing, we highlight a few here.  

One natural extension is to incorporate different definitions of the ecological distance between our populations. 
Just because two populations have no difference in their ecological variable state does not guarantee that there is not great heterogeneity in the distance between them.  For example, a pair of populations separated by the Grand Canyon might have nearly identical elevations, but the cost to migrants between them incurred by elevation may well be significant.  
One solution to this would be to enter a simple binary barrier variable, or
to calculate least-cost paths between populations, and use those distances in lieu of geographic distance.  
A more elegant solution would be to use ``isolation by resistance'' distances,
obtained by rasterizing landscapes and employing results relating mean passage rates of random walks in a heterogeneous environment to quantities from circuit theory in order to calculate the conductance (ease of migration) between nodes on that landscape \citep{McRaeBeier2007}.  This method has the advantage of integrating over all possible pathways between populations.
Currently, users must specify the resistance of landscape elements \emph{a priori}, but those resistance parameters could be incorporated into our parametric covariance function, and estimated along with the other parameters of our model in the same MCMC.  This approach carries great appeal, as it combines the conceptual rigor of accommodating multiple migration paths with the methodological rigor of statistically estimated, rather than user-specified, parameter values.  

Another extension is the further relaxation of the assumption of process homogeneity in decay of allelic covariance over geographic and ecological distance.  Specifically, the method currently requires that a single unit of pairwise ecological distance translate into the same extent of pairwise genetic differentiation between all population pairs.  This assumption is unlikely to be realistic in most empirical examples, especially if populations are locally adapted.  For example, individuals from populations adapted to high elevation may be able to migrate more easily over topography than individuals from populations adapted to low elevations.  Such heterogeneity could be accommodated by using different covariance functions for different, pre-specified population pairings.  

A final extension that could be integrated into this method is a model selection framework, in which models with and without an ecological distance variable, or with different combinations of ecological distance variables, can be rigorously compared. 
Because our method is implemented in a Bayesian framework, we could select between models by calculating Bayes factors (the ratio of the marginal likelihoods of the data under two competing hypotheses) \citep{Dickey1971, VerdinelliWasserman1995}.  
This approach would seem to offer the best of both worlds: robust parameter inference that accommodates uncertainty in addition to output that could be interpreted as definitive evidence for or against the association of an ecological variable of interest with genetic differentiation between populations.  

\subsection*{Conclusion}

In closing, we present a tool that can be useful in a wide variety of contexts,
allowing a description of the landscape as viewed by the movements of genetic material between populations.
We urge users to be cautious in their interpretation of results generated with this model.  A correlation between genetic differentiation and an ecological distance variable does not guarantee a causal relationship, especially because unmeasured ecological variables may be highly correlated with those included in an analysis.  In addition, evidence of a correlation between genetic differentiation and an ecological variable may not be evidence of local adaptation or selection against migrants, as both neutral and selective forces can give rise to an association between genetic divergence and ecological distance.  

Finally, we are making this method available online at \textit{genescape.org}, and we hope that users elaborate on the framework we present here to derive new models that are better able to describe empirical patterns of isolation by distance --- both geographic and ecological.

\subsection*{Acknowledgements}
We thank Yaniv Brandvain, Marjorie Weber, Luke Mahler, Will Wetzel, B. Moore and the Coop lab for their counsel, Jeff Ross-Ibarra and Torsten G\"{u}nther for their help with empirical datasets,  J. Novembre and D. Davison for their code, and Jon Wilkins and two anonymous reviewers for their comments on previous drafts.  
This material is based upon work supported by the National Science Foundation under Grant No. 1262645 (PR and GC), NSF GRFP No. 1148897 (GB), a NIH Ruth L. Kirschstein NRSA fellowship F32GM096686 (PR), and a Sloan Foundation fellowship (GC). 

\clearpage
\section*{Appendix}
\subsection*{Priors}
We denote a gamma distribution with given shape and rate parameters as $\Gamma(\mbox{shape},\mbox{rate})$, a normal distribution with given mean and variance parameters as $N(\mbox{mean},\mbox{variance})$, an exponential distribution with given rate parameter $\Exp(\mbox{rate})$, and a uniform distribution between given upper and lower boundaries as $U(\mbox{lower},\mbox{upper})$.  The priors specified on the parameters of this model are: $\alpha_{0} \sim \Gamma(0.001,0.001)$; $\alpha_{D} \sim \Exp(1)$; $\alpha_{E} \sim \Exp(1)$; $\alpha_{2} \sim U(0.1,2)$; and $\mu_{\ell} \sim N(0,1/\beta)$, with a hyper-prior $\beta \sim \Gamma(0.001,0.001)$.

The priors on $\alpha_D$ and $\alpha_E$ were chosen to reflect the assumption that there is some, and potentially very great, effect of isolation by geography and ecology.  The priors on $\alpha_2$, $\alpha_0$, and $\beta$ were the same as those used by \citep{Wasser2004}, and, in the case of the latter two (on $\beta$ and $\alpha_0$), were chosen because they were conjugate to the likelihood, so their parameters could therefore be updated by a Gibbs sampling step.  

In early implementations of our method, we experimented with uniform priors on  $\alpha_D$ and $\alpha_E$ (U(0,4)), as used by \citet{Wasser2004} (although they did not have a parameter analogous to $\alpha_E$).  We replaced these uniform priors with exponentials to reflect the fact that we have no prior belief that there should be any upper bound to the effects geographic or ecological distance may have on genetic differentiation.  In practice, we found that for all simulated and empirical datasets tested, there was sufficient information in the data for the likelihood function to swamp the effect of the priors --- whether uniform or exponential --- on $\alpha_D$ and $\alpha_E$.    

However, in all analyses, we encourage users to visualize the marginal distributions of each parameter at the end of a run and compare it to its prior.  If the marginal distribution looks exactly like the prior, there may be insufficient information in the data to parameterize the model effectively, and the prior may be having an unduly large impact on analysis.  If the marginal distribution for a parameter shows that it is ``piling up" against its prior's hard bound (e.g., the marginal distribution on $\alpha_E$ has a median of 1e-3, close to its hard bound at 0), that may suggest that the current form of the prior is not describing the natural distribution of the parameter for that particular dataset well (e.g., $\alpha_E$ ``wants" to be zero, but the prior is constraining it).  In both cases (the marginal posterior and the prior have significant overlap; the prior is exhibiting an edge effect), we suggest that the user experiment with different priors and/or model parameterizations to see what effect they are having on inference.

\subsection*{MCMC}
Our MCMC scheme proceeds as follows.  The chain is initiated at maximum likelihood estimates (MLEs) for $\theta$ and $\mu$, and, for  $\alpha_{0},\alpha_{D},\alpha_{E},$ and $\alpha_{2}$, at values drawn randomly from their priors.  The multiplicative inverse of the empirical variance of the MLEs of $\mu$ is used as the initial value of $\beta$.

In each generation one of $\{\mu,\beta,\theta,\alpha_{0},\alpha_{D},\alpha_{E},\alpha_{2}\}$ is selected at random to be updated. 

The priors on $\beta$ and $\alpha_{0}$ are conjugate to their marginal posteriors, and each is updated via a Gibbs sampling step.  The updated value of $\beta$ given the current $\mu$ is drawn from 
\begin{equation}
\beta \; | \; \mu_{1},\cdots,\mu_{L}  \sim \Gamma \left(0.001+\frac{L}{2},~0.001+\frac{1}{2}\sum\limits_{\ell=1}^{L}\mu_{\ell}^2 \right),
\end{equation}
and the updated value of $\alpha_{0}$ conditional on the current set of $\theta$ is drawn from 
\begin{equation}
\alpha_0 \; | \; \theta_{1},\cdots,\theta_{L} \sim \Gamma \left(1+\frac{Lk}{2}, ~1+\frac{1}{2}\sum\limits_{\ell=1}^{L}\theta_{\ell,k}\chi^{-1}\theta_{\ell,k}^{T} \right),
\end{equation}
where $k$ is the number of populations sampled, $L$ is the number of loci sequenced, and $\chi = \alpha_{0}\Omega = \exp{(-(\alpha_{D}D_{i,j}+\alpha_{E}E_{i,j})^{\alpha_{2}}})$.

The remaining parameters are updated by a Metropolis-Hastings step;
here we describe the proposal mechanisms.
The proposed updates to $\theta$ do not affect each other, and so are accepted or rejected independently.  Following Wasser \textit{et al.}~(2004) (derived from \citep{ChristensenWaagepetersen2002, Moller1998}), the proposal is chosen as $\theta_{\ell}^{'} = \theta_{\ell} +  R_{\ell}Z$, where $R$ is a vector of normally distributed random variables with mean zero and small variance (controlled by the scale of the tuning parameter on $\theta$) and $Z$ is the Cholesky decomposition of $\Omega$ (so that $ZZ^{T} = \Omega$).  Under this proposal mechanism, proposed updates to $\theta_{\ell}$ tend to stay within the region of high posterior probability, so that more updates are accepted and mixing is improved relative to a scheme in which the $\theta$ in each population were updated individually.  

Updates to $\alpha_{D}$, $\alpha_{E}$, and $\alpha_{2}$ are accomplished via a random-walk sampler (adding a normally distributed random variable with mean zero and small variance to the current value) \citep{Gilks1996}.  Updates to elements of $\mu_{\ell}$ are also accomplished via a random-walk sampler, and again the updates to each locus are accepted or rejected independently.  

In the overdispersion model, initial values of $\Phi_{k}$ are drawn from the prior for each population.  Updates are proposed one population at a time via a random-walk step, and are accepted or rejected independently.  

Well-suited values of tuning parameters (variances in the proposal distributions for $\mu,\theta,\alpha_{D},\alpha_{E}$, and $\alpha_{2}$)
and the number of generations required to accurately describe the joint posterior will vary from dataset to dataset, and so may require tuning.

\newpage
\bibliography{bedassle}

\clearpage

\section*{Supplemental material}
\renewcommand{\thefigure}{S\arabic{figure}}
\setcounter{figure}{0}
\renewcommand{\thetable}{S\arabic{table}}
\setcounter{table}{0}

\begin{figure}[ht!]
\begin{center}
  \includegraphics[width=6in,height=4in]{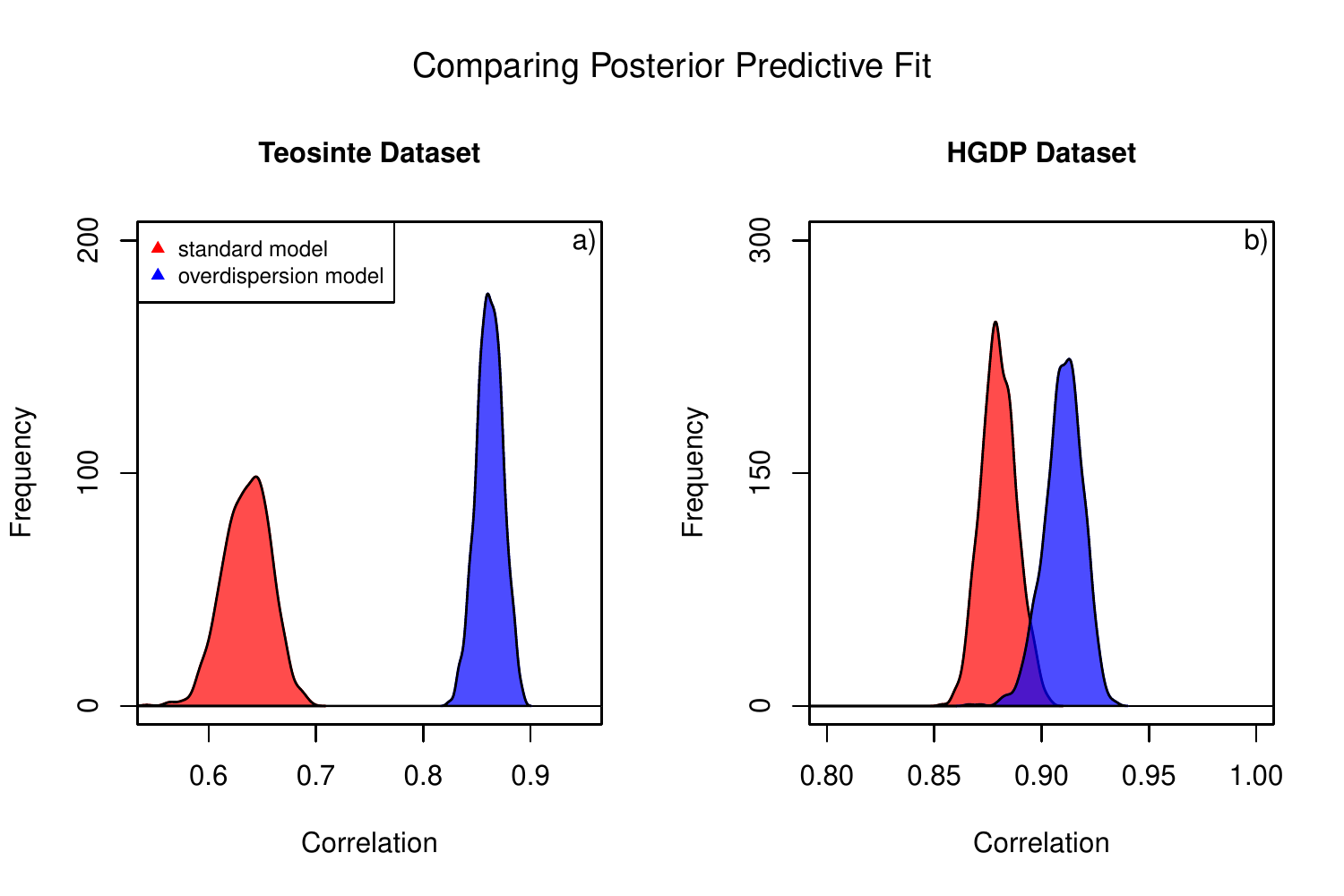}
 \caption{
 		\textmd{Distribution of Pearson's correlations between each posterior predictive simulated dataset and the observed data, highlighting the improved fit of the overdispersion model to describe:}
	\bf{a)}
 		\textmd{the teosinte dataset;}
	\bf{b)}
 		\textmd{the HGDP dataset.}
 \label{sfig:pps.corr}
  }
\end{center}
\end{figure}

\begin{figure}[ht!]
\begin{center}
  \includegraphics[width=6in,height=4.29in]{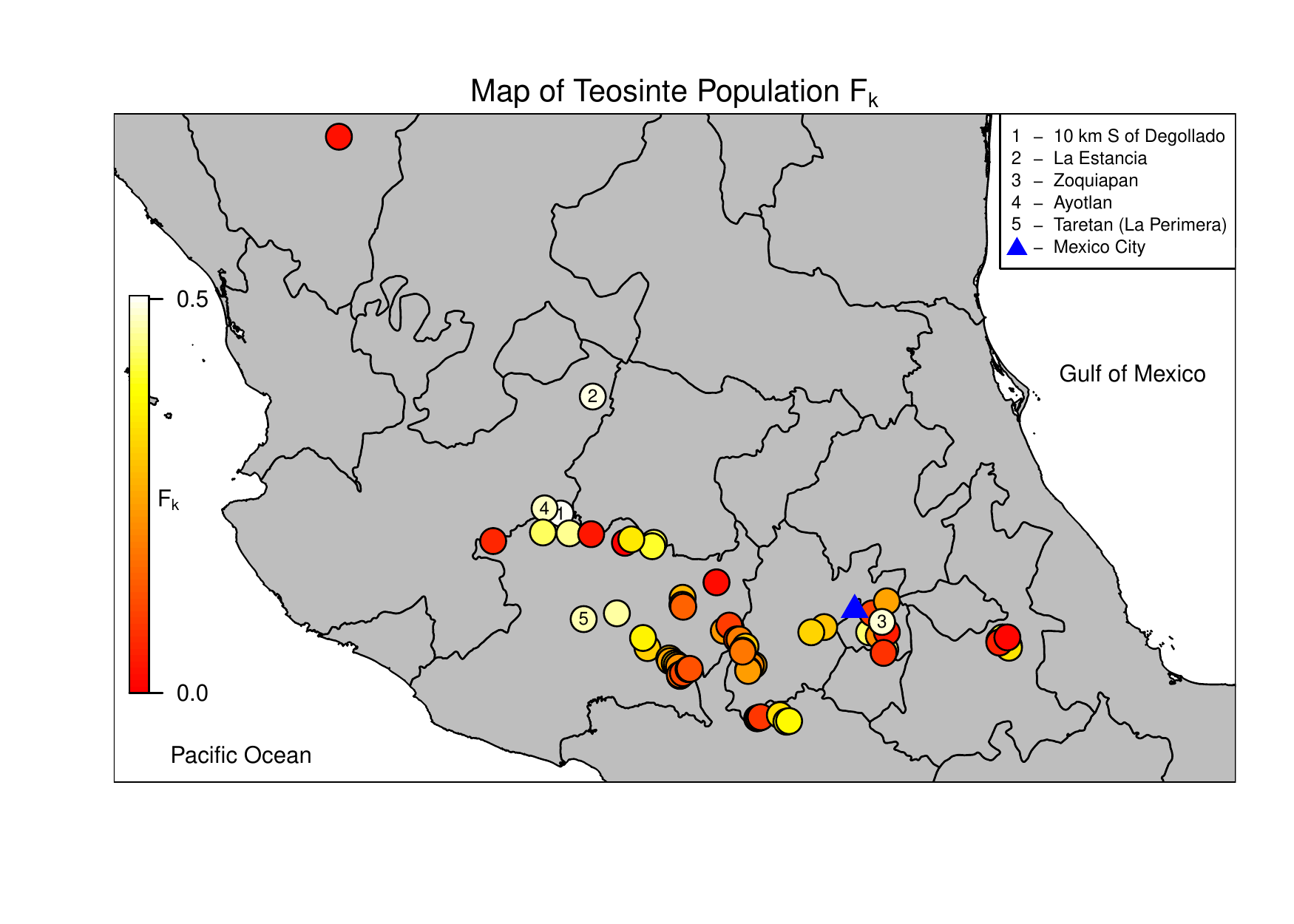}
 \caption{
\textmd{Map of teosinte populations sampled, colored by their median estimated population-specific overdispersion parameter, $F_{k}$.  The five populations with the highest values are noted.}
 \label{sfig:zea_Fk_map}
  }
\end{center}
\end{figure}

\begin{figure}[ht!]
\begin{center}
  \includegraphics[width=6in,height=3.4in]{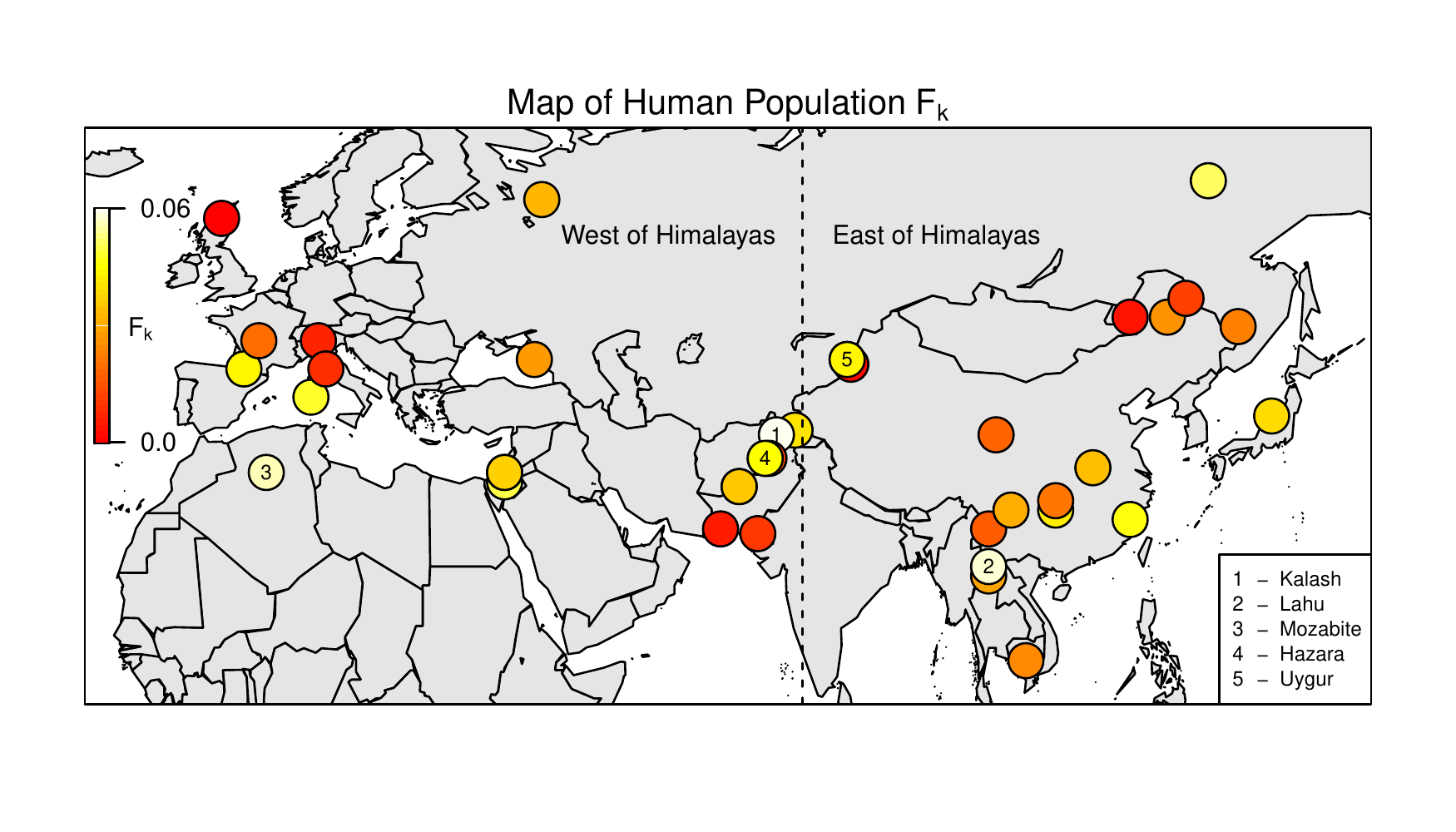}
 \caption{
\textmd{Map of human populations included in the analysis, colored by their median estimated population-specific overdispersion parameter, $F_{k}$.  The five populations with the highest values are noted.  
The dashed line denotes the line of longitude used to delimit the Himalayas.}
 \label{sfig:him_Fk_map}
  }
\end{center}
\end{figure}

\begin{figure}[ht!]
\begin{center}
  \includegraphics[width=6in,height=4in]{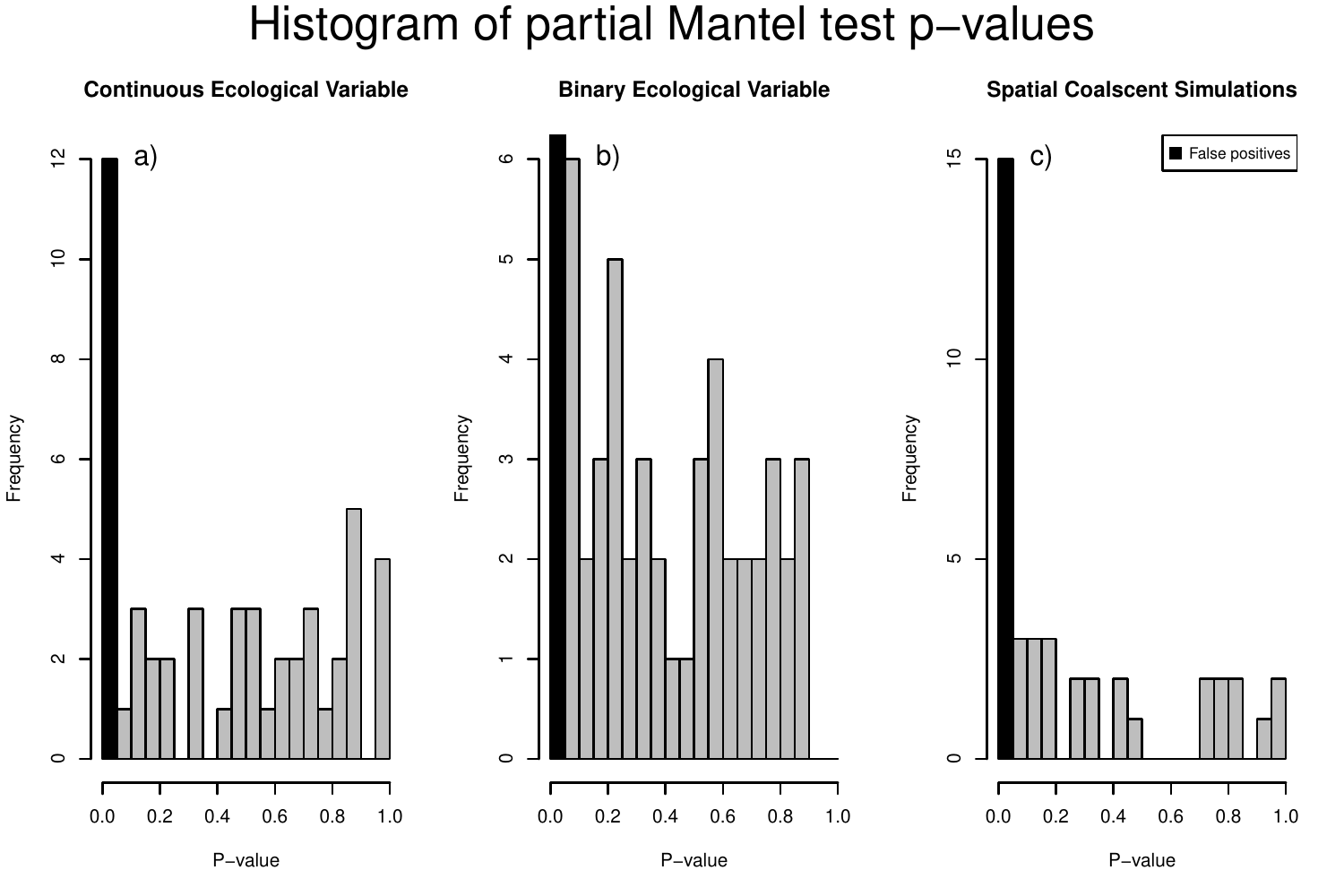}
 \caption{
 		\textmd{Histograms of p-values produced by the partial Mantel test (with 1,000,000 permutations) on the 140 datasets for which the true contribution of ecological distance to genetic differentiation was zero. The black column indicates the type I error rate with a significance level of p=0.05 in:}
	\bf{a)}
 		\textmd{the datasets with a continuous ecological distance variable;}
	\bf{b)}
 		\textmd{the datasets with a binary ecological distance variable.}
	\bf{c)}
 		\textmd{the datasets simulated under the spatial coalescent with a barrier that had no effect on genetic differentiation.}		
 \label{sfig:Pmantel}
  }
\end{center}
\end{figure}

\begin{figure}[ht!]
\begin{center}
  \includegraphics[width=5.25in,height=7in]{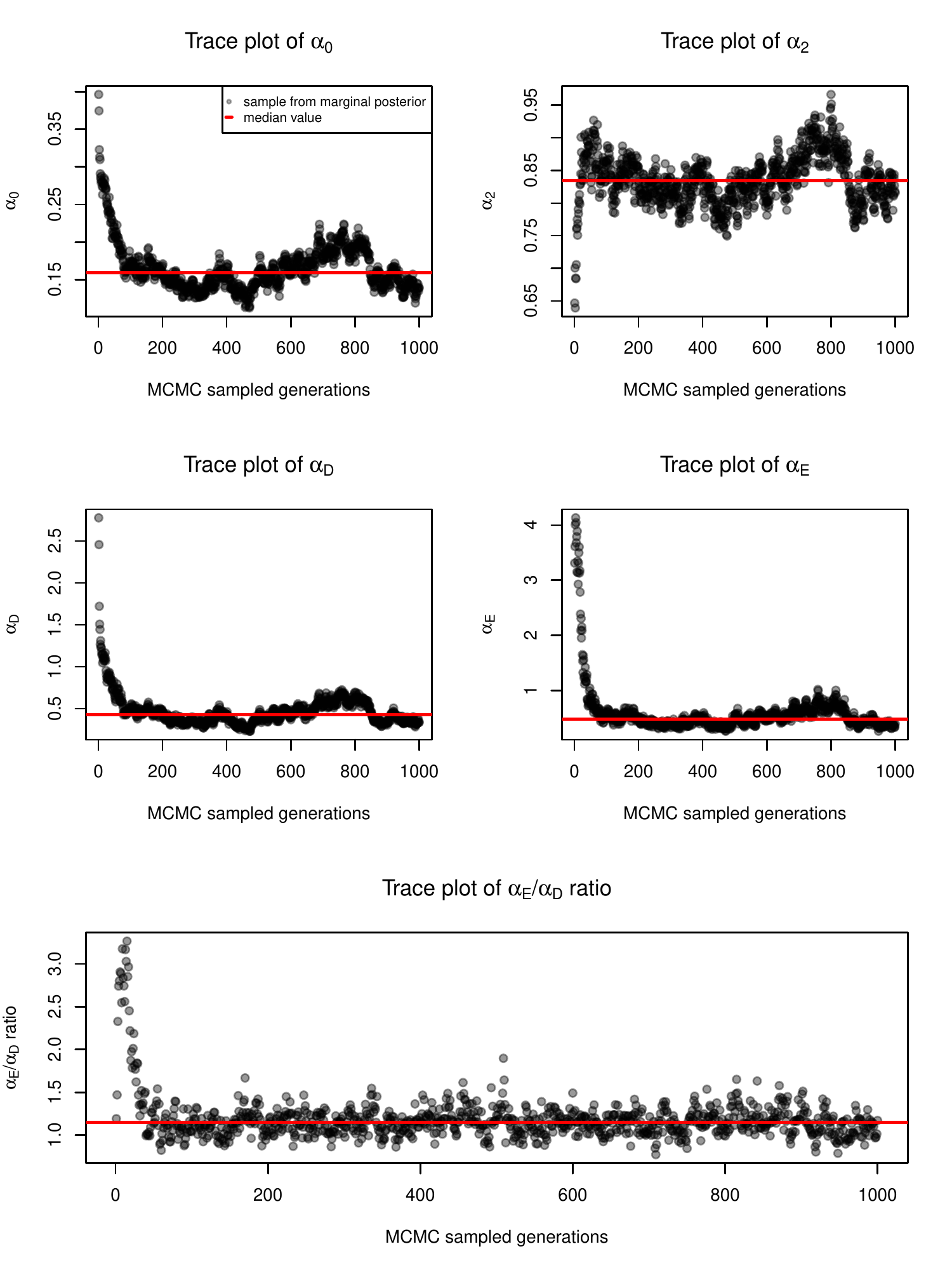}
 \caption{
 		\textmd{Trace plots of the $\alpha$ parameters of the covariance matrix $\Omega$.  Note the partial non-identifiability of the separate $\alpha$ parameters compared to the stability of the joint parameter, the $\alpha_E:\alpha_D$ ratio.}
 \label{sfig:trace_plots}
 }
\end{center}
\end{figure}

\begin{figure}[ht!]
\begin{center}
  \includegraphics[width=5.25in,height=7in]{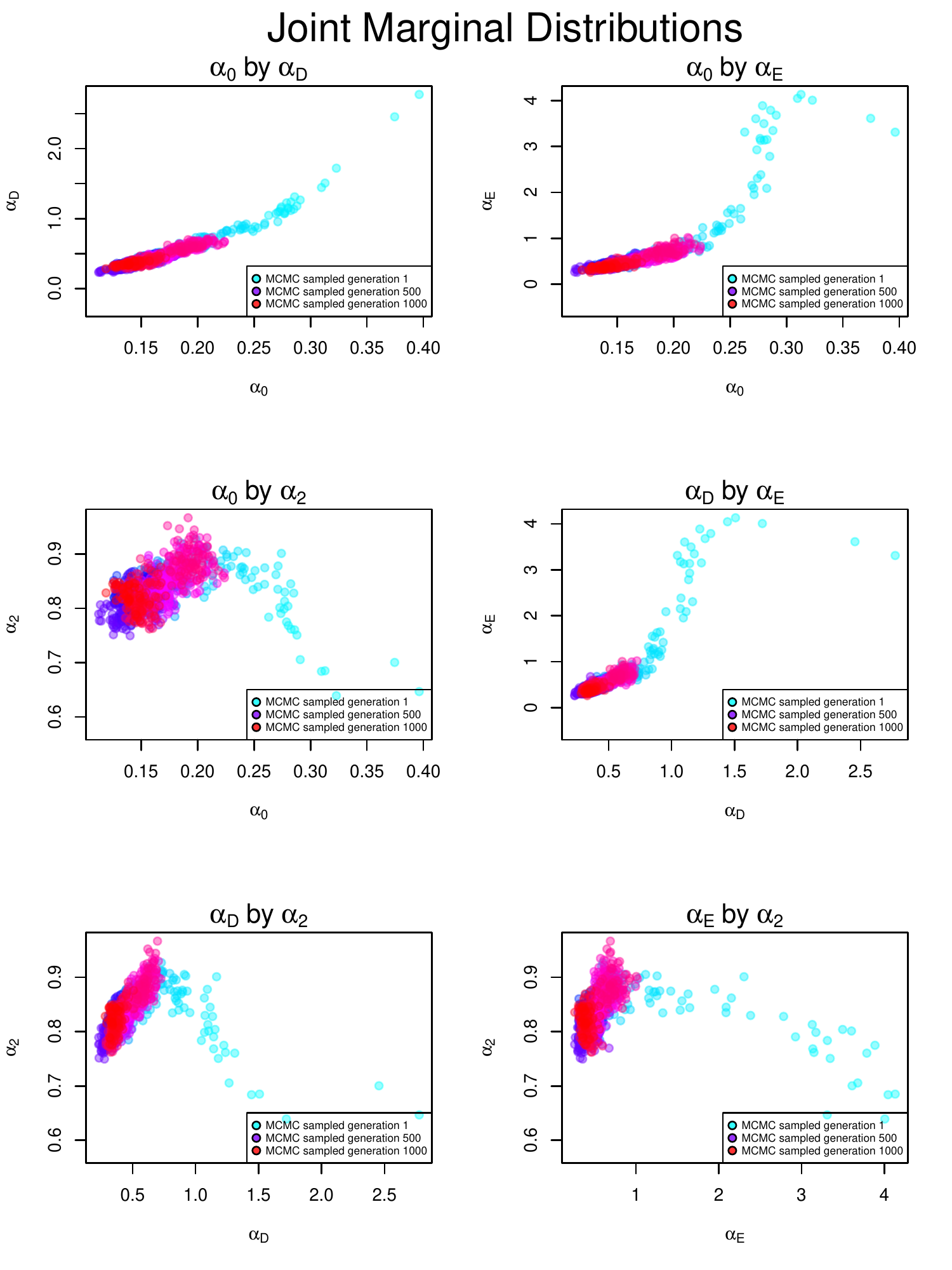}
 \caption{
 		\textmd{Joint marginal plots of the $\alpha$ parameters of the covariance matrix $\Omega$, colored by the MCMC  generation in which they were sampled.}
 \label{sfig:joint_marginals}
  }
\end{center}
\end{figure}

\begin{figure}[ht!]
\begin{center}
  \includegraphics[width=5.25in,height=7in]{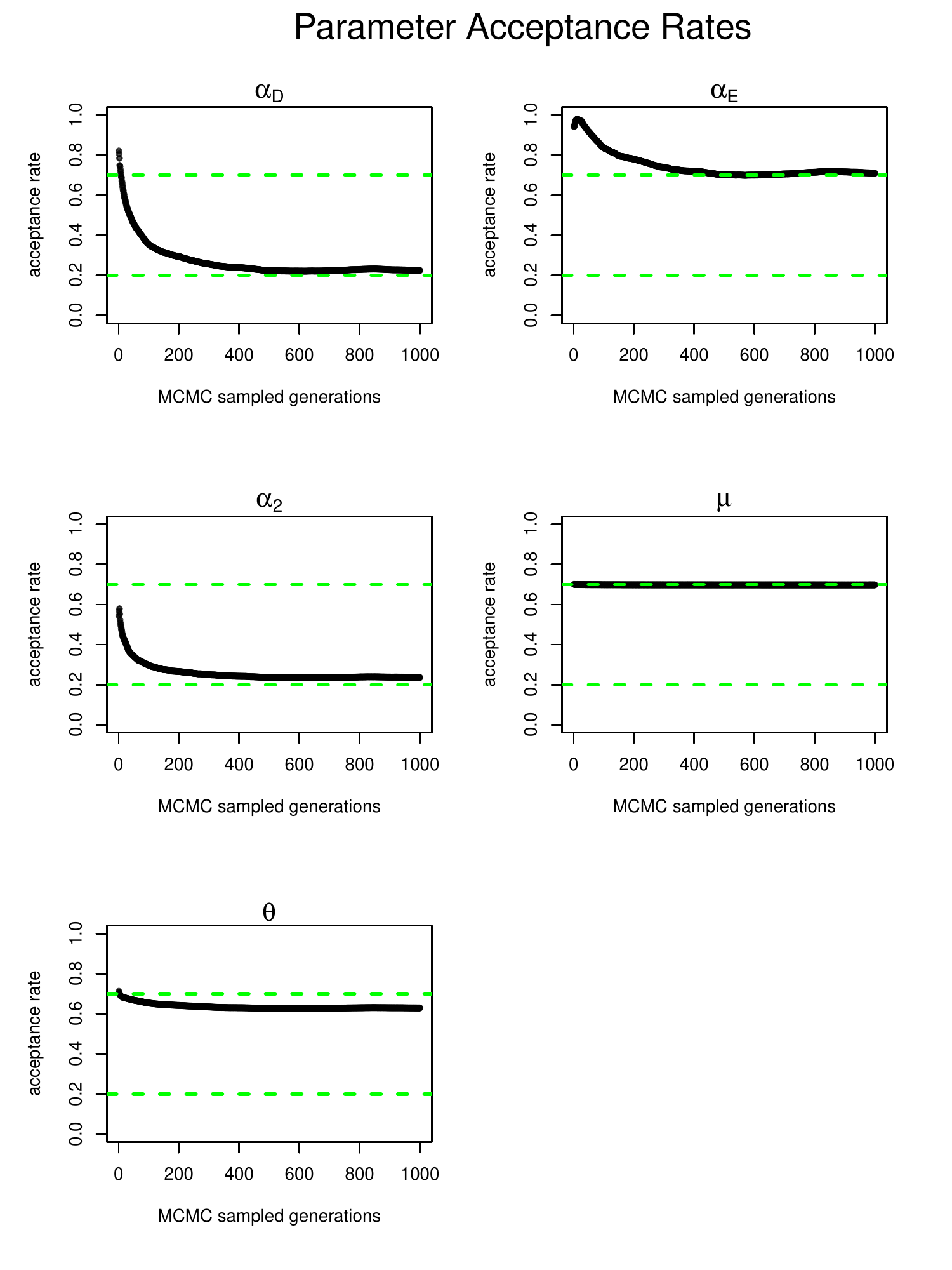}
 \caption{
 		\textmd{Acceptance rates for the parameters of the model that are updated with random-walk samplers, plotted over the duration of an individual MCMC run.  Dashed green lines indicate the bounds of acceptance rates that indicate optimal mixing: 20\%-70\%.}
 \label{sfig:acceptance_rates}
  }
\end{center}
\end{figure}

\begin{figure}[ht!]
\begin{center}
  \includegraphics[width=6in,height=3.4in]{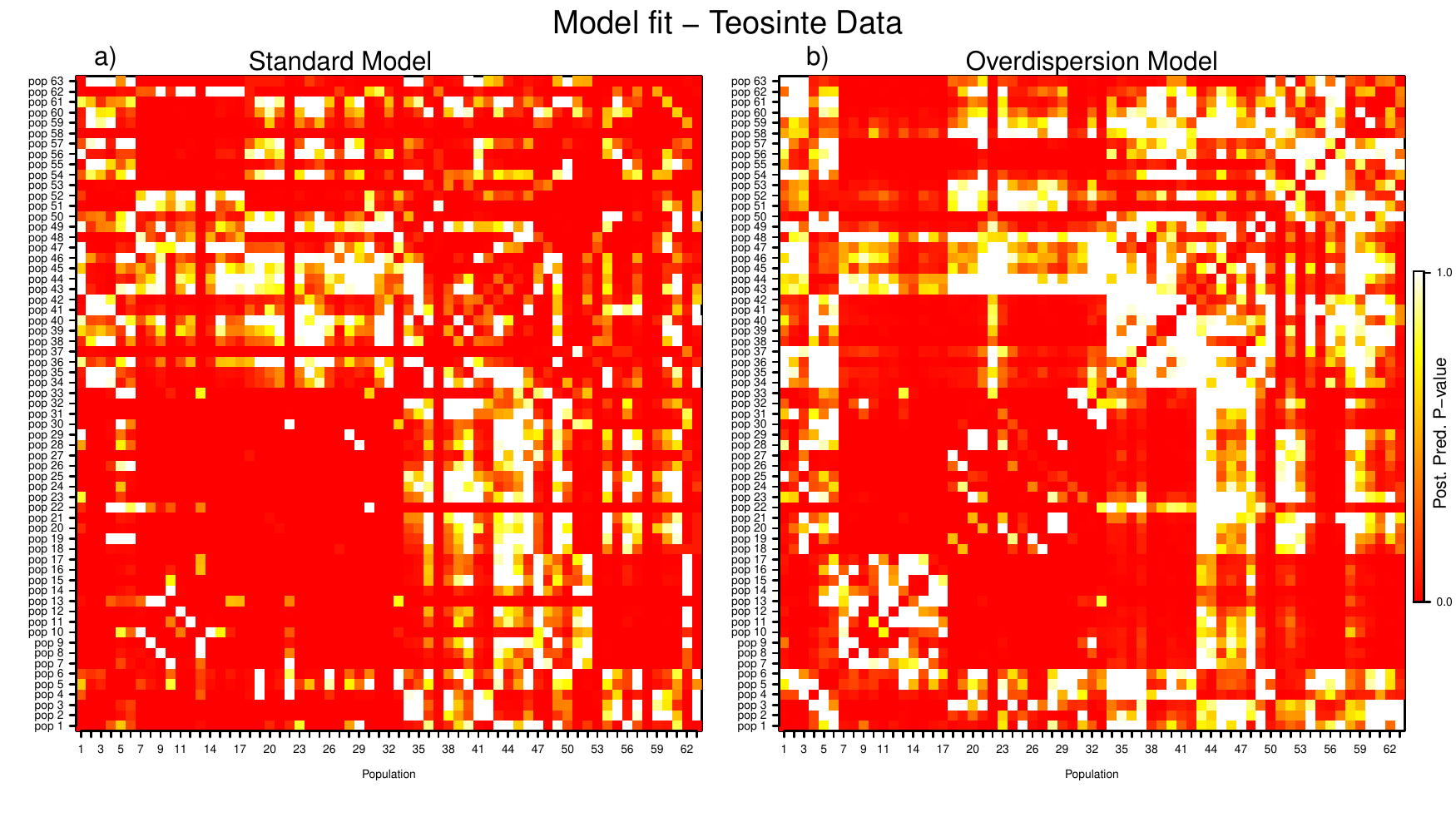}
 \caption{
		\textmd{Heatmapped matrices showing the performance of the model at all pairwise population comparisons.  The posterior predictive p-value was defined as 
		$1-2 \times |0.5-ecdf(F_{ST_{obs}})|$, in which $ecdf(F_{ST_{obs}})$ is the empirical cumulative probability of the observed $F_{ST}$ between two populations from a distribution defined by the posterior predictive sample for that population comparison, representing the p-value of a two-tailed t-test.  Higher p-values indicate better model fit.  Populations are enumerated on the margins, and may be referenced in SuppMat Table 1.}
	\bf{a)}
 		\textmd{The standard model.}
	\bf{b)}
 		\textmd{The overdispersion model.}
 \label{sfig:zea.pps.pval}
  }
\end{center}
\end{figure}

\begin{figure}[ht!]
\begin{center}
  \includegraphics[width=6in,height=3.4in]{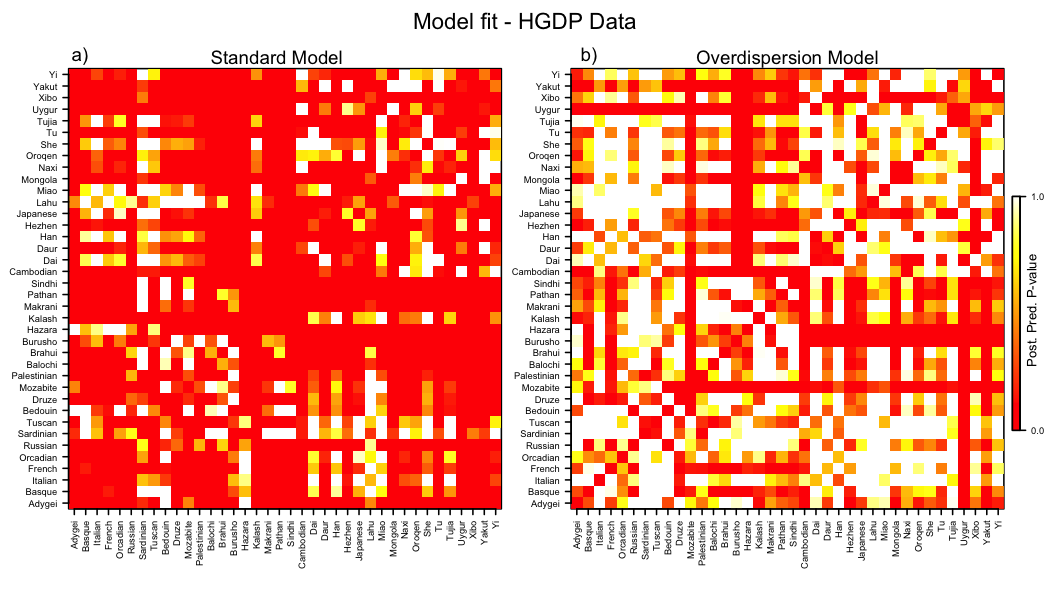}
 \caption{
\textmd{Heatmapped matrices indicating the performance of the model at all pairwise population comparisons.  The posterior predictive p-value was defined as 
		$1-2 \times |0.5-ecdf(F_{ST_{obs}})|$, in which $ecdf(F_{ST_{obs}})$ is the empirical cumulative probability of the observed $F_{ST}$ between two populations from a distribution defined by the posterior predictive sample for that population comparison, representing the p-value of a two-tailed t-test.  Higher p-values indicate better model fit.  Populations are enumerated on the margins, and may be referenced in SuppMat Table 2.}
	\bf{a)}
 		\textmd{The standard model.}
	\bf{b)}
 		\textmd{The overdispersion model.}
\label{sfig:him.pps.pval}
  }
\end{center}
\end{figure}

\begin{figure}[ht!]
\begin{center}
  \includegraphics[width=4.66in,height=2.66in]{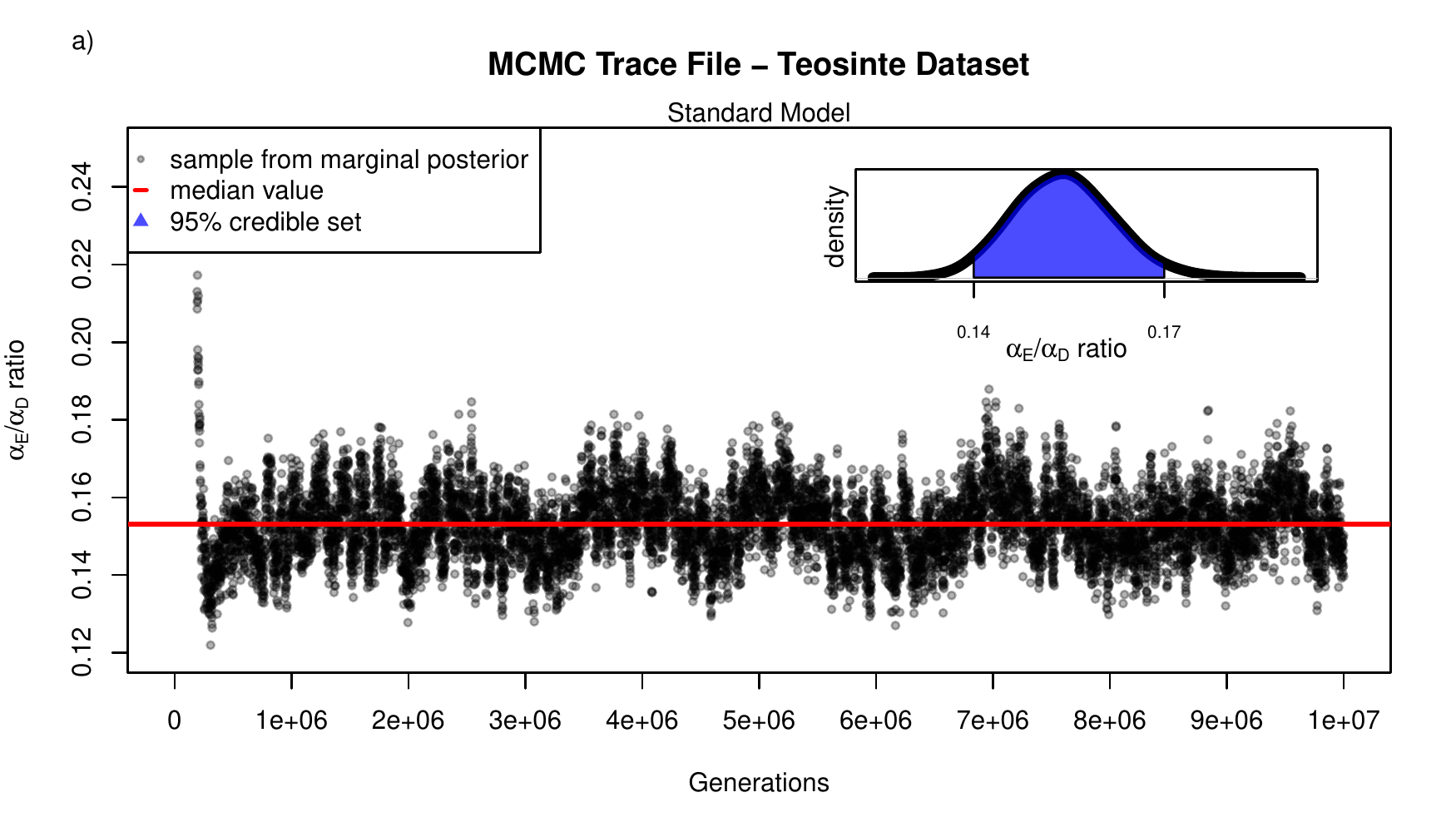}
  \includegraphics[width=4.66in,height=2.66in]{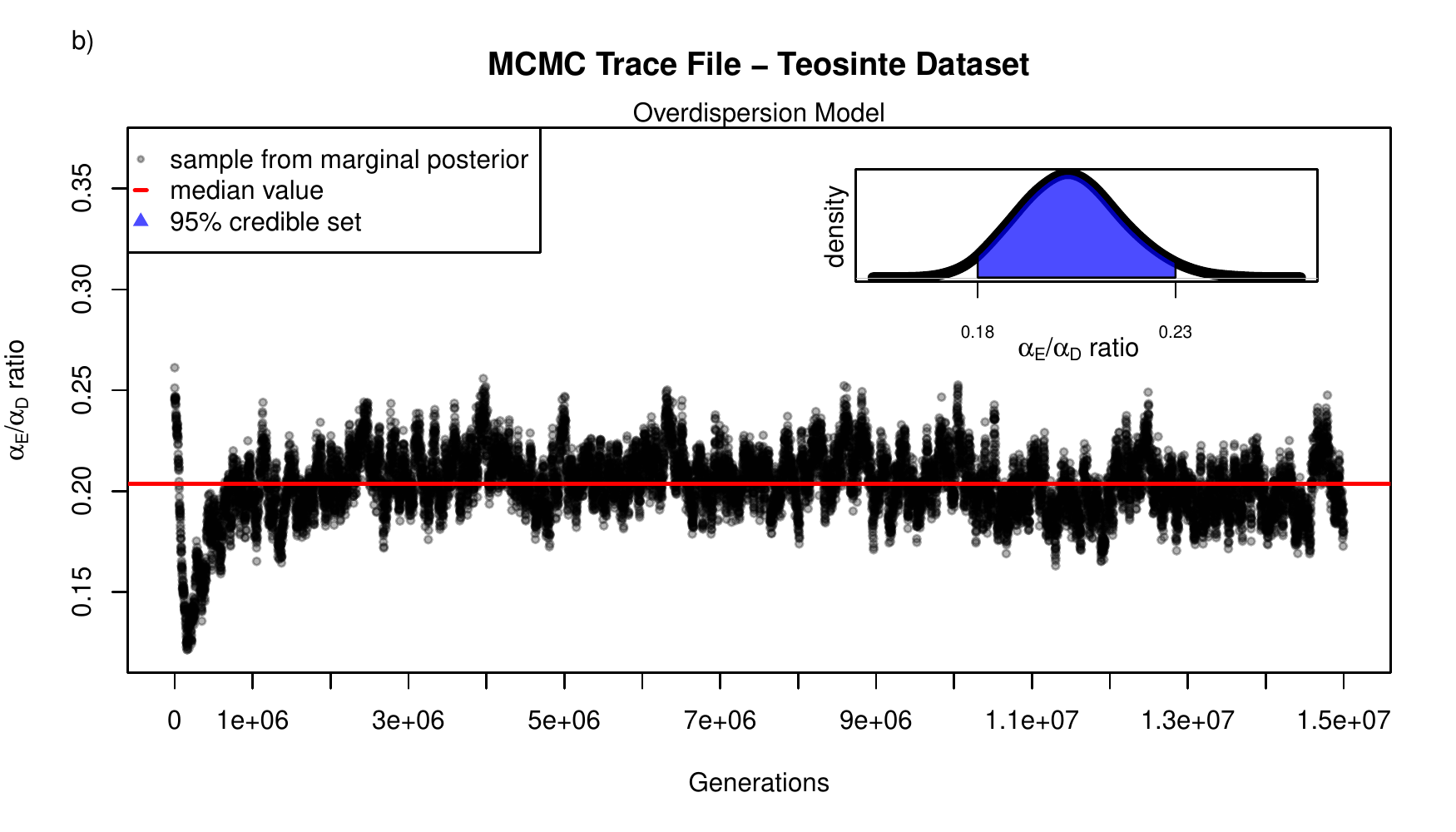}  
 \caption{
\textmd{Trace plots of the marginal posterior estimates for the $\alpha_{E} / \alpha_{D}$ ratio from MCMC analysis of the teosinte dataset.  Inset figures 
		give the marginal densities and 95\% credible set for the samples after a burn-in of 20\%}
	\bf{a)}
 		\textmd{The standard model.}
	\bf{b)}
 		\textmd{The overdispersion model.}
\label{sfig:zea_traceplot}
  }
\end{center}
\end{figure}

\begin{figure}[ht!]
\begin{center}
  \includegraphics[width=4.66in,height=2.66in]{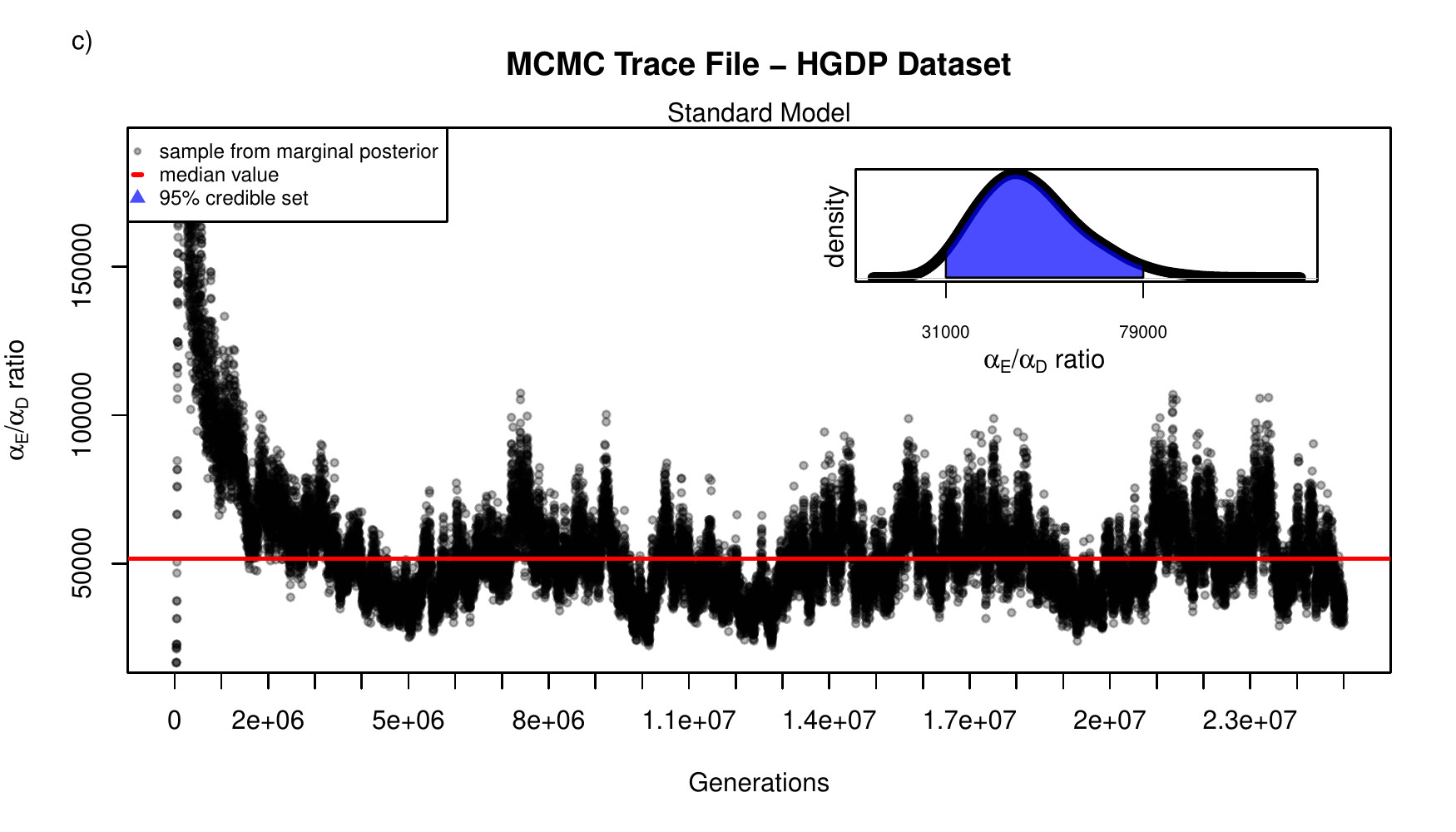}
  \includegraphics[width=4.66in,height=2.66in]{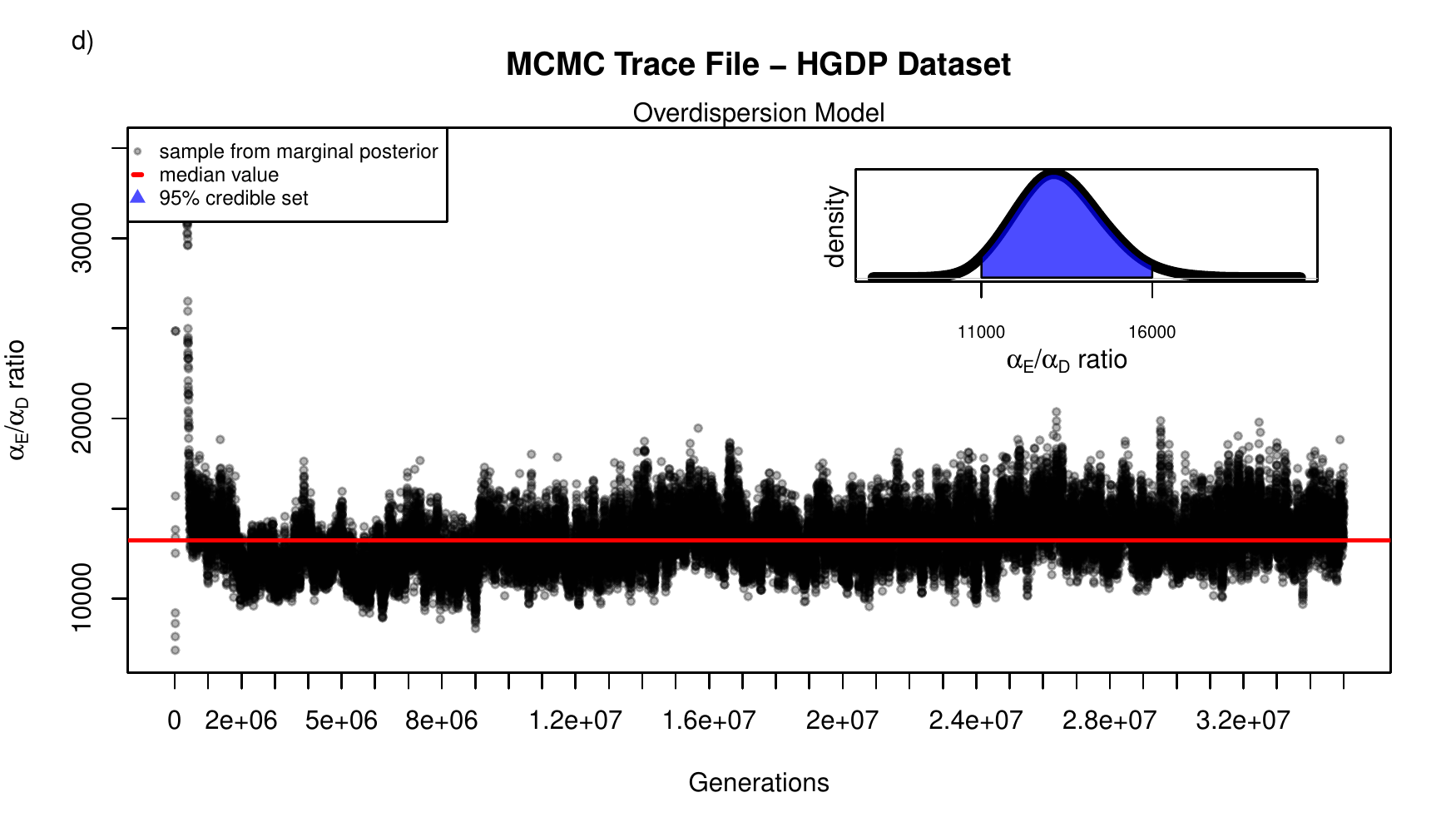}  
 \caption{
\textmd{Trace plots of the marginal posterior estimates for the $\alpha_{E} / \alpha_{D}$ ratio from MCMC analysis of the HGDP dataset.  Inset figures 
		give the marginal densities and 95\% credible set for the samples after a burn-in of 20\%}
	\textbf{a)}
 		The standard model.
	\textbf{b)}
 		The overdispersion model.
\label{sfig:him_traceplot}
  }
\end{center}
\end{figure}

\begin{table}
\begin{center}
\tiny{
    \begin{tabular}{r@{--}lllll}
  \hline
  & Population name (sample size) & Latitude & Longitude & Elevation & Subspecies \\ 
  \hline
1 & Km 1 El Crustel-Teloloapan (44) & 18.383 & 18.383 & 985 & parviglumis \\ 
2 & Amates Grandes (50) & 18.388 & 18.388 & 1110 & parviglumis \\ 
3 & Km 3 Amates Grandes-Teloloapan (48) & 18.394 & 18.394 & 1210 & parviglumis \\ 
4 & Km 72 Iguala-Arcelia (Km Alcholoa-Arcelia) (56) & 18.414 & 18.414 & 1506 & parviglumis \\ 
5 & Rinc\'on del Sauce (56) & 18.35 & 18.35 & 1624 & parviglumis \\ 
6 & Ahuacatitl\'an (km 1.5 del entronque) (38) & 18.356 & 18.356 & 1528 & parviglumis \\ 
7 & Km 80 Huetamo-Villa Madero (50) & 19.063 & 19.063 & 832 & parviglumis \\ 
8 & Puerto de la Cruz (Km 119 Huetamo-V.Madero) (40) & 18.963 & 18.963 & 870 & parviglumis \\ 
9 & El Zapote (km 122 Huetamo-Caracuaro) (50) & 18.938 & 18.938 & 915 & parviglumis \\ 
10 & Puerto El Coyote (40) & 18.916 & 18.916 & 727 & parviglumis \\ 
11 & Km 135-136 Huetamo-Villa Madero (40) & 18.9 & 18.9 & 677 & parviglumis \\ 
12 & Cuirindalillo (km 142 Huetamo-Caracuaro) (42) & 18.883 & 18.883 & 697 & parviglumis \\ 
13 & Crucero Puertas de Chiripio (50) & 18.794 & 18.794 & 653 & parviglumis \\ 
14 & Quenchendio (km 151.5 Zit\'acuaro-Huetamo) (54) & 18.805 & 18.805 & 635 & parviglumis \\ 
15 & El Potrero (km 145.5 Zit\'acuaro-Huetamo) (40) & 18.82 & 18.82 & 654 & parviglumis \\ 
16 & La Crucita (km 135 Zit\'acuaro-Huetamo) (58) & 18.858 & 18.858 & 609 & parviglumis \\ 
17 & El Guayabo (km 132.5 Zit\'acuaro-Huetamo) (54) & 18.862 & 18.862 & 555 & parviglumis \\ 
18 & Km 107-108 Toluca-Altamirano (50) & 18.899 & 18.899 & 1422 & parviglumis \\ 
19 & Km 112 Toluca-Altamirano (46) & 18.895 & 18.895 & 1355 & parviglumis \\ 
20 & Km 119 Toluca-Altamirano (38) & 18.854 & 18.854 & 1015 & parviglumis \\ 
21 & Salitre-Monte de Dios (46) & 18.842 & 18.842 & 958 & parviglumis \\ 
22 & Taretan (La Perimera) (36) & 19.344 & 19.344 & 1170 & parviglumis \\ 
23 & Los Guajes (km 43 Zit\'acuaro-Huetamo) (54) & 19.231 & 19.231 & 985 & parviglumis \\ 
24 & 1 Km Norte de Santa Ana (54) & 19.281 & 19.281 & 1332 & parviglumis \\ 
25 & Km 8 Zuluapan-Tingambato (58) & 19.148 & 19.148 & 1178 & parviglumis \\ 
26 & Km 4 Zuluapan-Tingambato (60) & 19.146 & 19.146 & 1346 & parviglumis \\ 
27 & K2 Zacazonapan-Otzoloapan (56) & 19.079 & 19.079 & 1468 & parviglumis \\ 
28 & K22 Zacazonapan-Luvianos (EL Puente) (56) & 19.039 & 19.039 & 1085 & parviglumis \\ 
29 & Acatitl\'an-El Puente (50) & 19.029 & 19.029 & 1075 & parviglumis \\ 
30 & Queretanillo (56) & 19.551 & 19.551 & 1342 & parviglumis \\ 
31 & Km 33.5 Temascal-Huetamo (56) & 19.483 & 19.483 & 1100 & parviglumis \\ 
32 & Km 37 Temascal-Huetamo (40) & 19.464 & 19.464 & 1030 & parviglumis \\ 
33 & Casa Blanca (km 62 Huetamo-Villa Madero) (54) & 19.161 & 19.161 & 1268 & parviglumis \\ 
34 & San Antonio Tecomitl (4) & 19.217 & 19.217 & 2400 & mexicana \\ 
35 & Ozumba (4) & 19.05 & 19.05 & 2340 & mexicana \\ 
36 & Temamatla (6) & 19.183 & 19.183 & 2400 & mexicana \\ 
37 & Zoquiapan (4) & 19.317 & 19.317 & 2270 & mexicana \\ 
38 & Los Reyes La Paz (6) & 19.4 & 19.4 & 2200 & mexicana \\ 
39 & Miraflores (4) & 19.217 & 19.217 & 2200 & mexicana \\ 
40 & Tepetlixpa (4) & 19.017 & 19.017 & 2320 & mexicana \\ 
41 & El Pedregal (4) & 19.267 & 19.267 & 2500 & mexicana \\ 
42 & Mexicaltzingo (4) & 19.217 & 19.217 & 2600 & mexicana \\ 
43 & Santa Cruz (4) & 19.083 & 19.083 & 2425 & mexicana \\ 
44 & San Antonio (4) & 19.067 & 19.067 & 2440 & mexicana \\ 
45 & San Salvador (4) & 19.133 & 19.133 & 2425 & mexicana \\ 
46 & Tlachichuca (4) & 19.167 & 19.167 & 2355 & mexicana \\ 
47 & K3 San Salvador El Seco-Coatepec (4) & 19.117 & 19.117 & 2425 & mexicana \\ 
48 & San Nicolas B. Aires (4) & 19.167 & 19.167 & 2355 & mexicana \\ 
49 & San Felipe (4) & 19.517 & 19.517 & 2250 & mexicana \\ 
50 & 4 miles N of Hidalgo, Arroyo Zarco (4) & 19.7 & 19.7 & 2040 & mexicana \\ 
51 & 5-7 km SW Cojumatlan (4) & 20.1 & 20.1 & 1700 & mexicana \\ 
52 & Puente Gavilanes (4) & 24.017 & 24.017 & 1950 & mexicana \\ 
53 & La Estancia (4) & 21.5 & 21.5 & 1920 & mexicana \\ 
54 & Moroleon (4) & 20.083 & 20.083 & 2100 & mexicana \\ 
55 & Pinicuaro (8) & 20.05 & 20.05 & 2087.5 & mexicana \\ 
56 & Puruandiro (4) & 20.083 & 20.083 & 2000 & mexicana \\ 
57 & km 2 Puruandiro-Las Tortugas (4) & 20.117 & 20.117 & 1880 & mexicana \\ 
58 & 10 km S of Degollado (4) & 20.367 & 20.367 & 1625 & mexicana \\ 
59 & Ayotlan (4) & 20.417 & 20.417 & 1520 & mexicana \\ 
60 & Churitzio (8) & 20.175 & 20.175 & 1780 & mexicana \\ 
61 & El Salitre 1-2 km SE (4) & 20.183 & 20.183 & 1530 & mexicana \\ 
62 & Rancho El Tejocote (4) & 20.167 & 20.167 & 1750 & mexicana \\ 
63 & Villa Escalante (6) & 19.4 & 19.4 & 2320 & mexicana \\ 
   \hline
\end{tabular}
}
\label{tab:zea_popdata}
\end{center}
\caption{Metadata for populations used in the teosinte dataset.}
\end{table}

\begin{table}
\begin{center}
\tiny{
    \begin{tabular}{r@{--}lccc}
  \hline
  & Population name (sample size) & Latitude & Longitude & Side of the Himalayas \\ 
  \hline
1 & Adygei (30) & 44 & 39 & W \\ 
2 & Basque (36) & 43 & 0 & W \\ 
3 & Italian (20) & 46 & 10 & W \\ 
4 & French (52) & 46 & 2 & W \\ 
5 & Orcadian (28) & 59 & -3 & W \\ 
6 & Russian (46) & 61 & 40 & W \\ 
7 & Sardinian (46) & 40 & 9 & W \\ 
8 & Tuscan (10) & 43 & 11 & W \\ 
9 & Bedouin (86) & 31 & 35 & W \\ 
10 & Druze (78) & 32 & 35 & W \\ 
11 & Mozabite (50) & 32 & 3 & W \\ 
12 & Palestinian (88) & 32 & 35 & W \\ 
13 & Balochi (44) & 30.5 & 66.5 & W \\ 
14 & Brahui (46) & 30.5 & 66.5 & W \\ 
15 & Burusho (48) & 36.5 & 74 & W \\ 
16 & Hazara (40) & 33.5 & 70 & W \\ 
17 & Kalash (44) & 36 & 71.5 & W \\ 
18 & Makrani (48) & 26 & 64 & W \\ 
19 & Pathan (40) & 33.5 & 70.5 & W \\ 
20 & Sindhi (44) & 25.5 & 69 & W \\ 
21 & Cambodian (16) & 12 & 105 & E \\ 
22 & Dai (18) & 21 & 100 & E \\ 
23 & Daur (14) & 48.5 & 124 & E \\ 
24 & Han (64) & 32.5 & 114 & E \\ 
25 & Hezhen (16) & 47.5 & 133.5 & E \\ 
26 & Japanese (50) & 38 & 138 & E \\ 
27 & Lahu (12) & 22 & 100 & E \\ 
28 & Miao (10) & 28 & 109 & E \\ 
29 & Mongola (18) & 48.5 & 119 & E \\ 
30 & Naxi (14) & 26 & 100 & E \\ 
31 & Oroqen (16) & 50.5 & 126.5 & E \\ 
32 & She (18) & 27 & 119 & E \\ 
33 & Tu (18) & 36 & 101 & E \\ 
34 & Tujia (18) & 29 & 109 & E \\ 
35 & Uygur (18) & 44 & 81 & E \\ 
36 & Xibo (16) & 43.5 & 81.5 & E \\ 
37 & Yakut (46) & 63 & 129.5 & E \\ 
38 & Yi (18) & 28 & 103 & E \\ 
   \hline
\end{tabular}
}
\label{tab:hgdp_popdata}
\end{center}
\caption{Metadata for populations used from the HGDP dataset.}
\end{table}

\end{document}